\documentclass[preprint]{aastex}
\usepackage{graphics,graphicx,natbib}
\begin{document}
\title{A Dynamical Model for the Evolution of a Pulsar Wind Nebula
  inside a Non-Radiative Supernova Remnant} \author{Joseph
  D. Gelfand\altaffilmark{a}\altaffilmark{b}, Patrick
  O. Slane\altaffilmark{c}, Weiqun Zhang\altaffilmark{a}} 
\altaffiltext{a}{New York University}
\altaffiltext{b}{NSF Astronomy and Astrophysics Postdoctoral Fellow}
\altaffiltext{c}{Harvard-Smithsonian Center for Astrophysics}

\begin{abstract}
A pulsar wind nebula inside a supernova remnant provides a unique
insight into the properties of the central neutron star, the
relativistic wind powered by its loss of rotational energy, its
progenitor supernova, and the surrounding environment.  In this paper,
we present a new semi-analytic model for the evolution of such a
pulsar wind nebula.  This model couples the dynamical and radiative
evolution of the pulsar wind nebulae, traces the evolution of the
pulsar wind nebulae throughout the lifetime of the supernova remnant
produced by the progenitor explosion, and predicts both the dynamical
(e.g. radius and expansion velocity) and radiative (radio to TeV
$\gamma$-ray spectrum) properties of the pulsar wind nebula during
this period.  As a result, it is uniquely qualified for using the
observed properties of a pulsar wind nebula in order to constrain the
physical characteristics of the neutron star, pulsar wind, progenitor
supernova, and surrounding interstellar medium.  We also discuss the
expected evolution for a particular set of these parameters, and show
that it reproduced the large spectral break observed in radio and
X-ray observations of many young pulsar wind nebulae, and the low
break frequency, low radio luminosity and high TeV $\gamma$-ray
luminosity, and high magnetization observed for several older pulsar
wind nebulae.  The predicted spectrum of this pulsar wind nebula also
contains spectral features during different phases of its evolution
detectable with new radio and $\gamma$-ray observing facilities such as
the Extended Very Large Array and the {\it Fermi Gamma-ray Space
Telescope}.  Finally, this model has implications for determining
if pulsar wind nebulae can inject a sufficient number of energetic
electrons and positrons into the surrounding interstellar medium to
explain the recent measurements of the cosmic ray positron fraction by
{\it PAMELA} and the cosmic ray lepton spectrum by ATIC and HESS.
\end{abstract}
\keywords{ISM: supernova remnants, stars: pulsars: general}

\section{Introduction}
\label{intro}
The gravitational collapse of the core of a massive star into a
neutron star (e.g. \citealt{baade34}) releases enough energy to power
a supernova explosion (e.g. \citealt{zwicky38}).  The resultant
neutron star is born spinning, and thought to lose its rotational
energy through a ultra-relativistic magnetic and particle outflow
commonly referred to as a pulsar wind (e.g. \citealt{goldreich69,
kennel84}).  The interaction between the pulsar wind and neutron
star's environment creates an object called a pulsar wind nebula
(PWN).  Initially ($\la10^5$ years after the supernova explosion), the
neutron star and its PWN are inside the supernova remnant (SNR)
created by the expansion of the supernova ejecta into the surrounding
interstellar medium (ISM).  Previous work (e.g. \citealt{pacini73,
reynolds84}) has demonstrated that, during this period, the properties
of the PWN depends on the physical properties of the central neutron
star (e.g. its space velocity, initial period, surface magnetic field
strength, spin-down timescale, and braking index), the pulsar wind
(e.g. its magnetization and energy spectrum of particles injected into
the PWN), the progenitor supernova explosion (e.g. the mass and
initial kinetic energy of the ejecta), and the surrounding ISM
(e.g. the density profile).  Measuring these quantities is important
for understanding the underlying physical mechanism behind
core-collapse supernova, the formation of neutron stars in these
explosions, and the evolution and properties of the progenitor star.
Many of these quantities are extremely difficult to measuring directly
for most neutron stars / SNRs, but possible to infer indirectly using
observations of PWNe.

Using the observed properties of a PWN inside a SNR to determine the
properties of the neutron star, pulsar wind, progenitor supernova, and
ISM requires understanding its evolution.  As summarized in a recent
review by \citet{gaensler06}, this is extremely complicated due to the
rapid evolution of both the SNR and central neutron star.  Analytical
(e.g. \citealt{ostriker71, pacini73, reynolds84}) and numerical
simulations (e.g. \citealt{bucciantini03, vdswaluw04}) of the
evolution of a PWN inside a SNR identify three important evolutionary
phases (e.g. \citealt{gelfand07,gaensler06}): an initial
free-expansion, the eventual collision between the PWN and SNR reverse
shock which causes the PWN to contract and re-expand inside the SNR --
eventually stripping the neutron star of its PWN, and a two PWNe phase
inside the SNR - a ``relic'' PWN containing the particles injected
into the neutron star at earlier times and a ``new'' PWN composed of
particles injected by the neutron star after it leaves the ``relic''
PWN.  Not surprisingly, the observable properties of a PWN vary
significantly during this evolution, and are highly sensitive to the
physical characteristics of the central neutron star, progenitor
supernova, pulsar wind, and ISM listed above.  Therefore, in order to
use the observed properties of a PWN to constrain these quantities, it
is necessary to have a model for the evolution of a PWN inside a SNR
which takes into account the relevant physics of all of these
components.

Many such models exist in literature (e.g. \citealt{blondin01,
bucciantini03, bucciantini04, chevalier92, chevalier05, delzanna04,
gelfand07, jun98, kennel84, kennel84b, ostriker71, pacini73, rees74,
reynolds84, vdswaluw01b, vdswaluw03, vdswaluw04, venter06, volpi08}
and references therein).  While most reproduce the general
evolutionary sequence outlined above, they differ significantly in the
details, e.g.  their treatment of the evolution of the PWN's magnetic
field, the injection of the pulsar wind into the PWN.  These
differences have significant consequences for the predicted evolution
of the observable properties of a PWN (such as its size and broadband
spectrum).  Additionally, most models predict either the dynamical
properties (e.g. size and expansion velocity; \citealt{vdswaluw01}) or
spectral properties (e.g. broadband spectrum; \citealt{volpi08}) of
the PWN but not both.  In this paper, we present a new semi-analytic
model for the evolution of a PWN which predicts both the dynamical and
radiative properties of a PWN inside a non-radiative SNR through the
entire evolutionary sequence described above.  This is important
because there are many systems of interest where the PWN likely has
collided with the SNR reverse shock (e.g. G328.4+0.2,
\citealt{gelfand07}; G327.1--1.1, \citealt{temim09}; Vela X
\citealt{lamassa08}).  This model also self-consistently couples the
dynamical and radiative evolution of the PWN, including the evolution
of the PWN's magnetic field, and calculates the broadband (radio - TeV
$\gamma$-ray evolution) of the PWN.  As a result, it is is well suited
for both examining the effect of different supernova, neutron star,
pulsar wind, and ISM properties on the evolution of the resulting PWN,
and for using the observed properties of a PWN to determine the
physical properties of the progenitor supernova, central neutron star,
and surrounding ISM.

This paper is structured as follows.  In \S\ref{theory}, we describe
the physics underlying our model and its implementation.  In
\S\ref{performance}, we present and discuss the predicted evolution of
a PWN for a particular set of input parameters.  Finally, in
\S\ref{discussion}, we discuss the implication of these result and
potential applications of this model.

\section{Model Description and Implementation}
\label{theory}
In this Section, we describe the underlying physics of this model for
the evolution of a PWN inside a SNR (\S\ref{theory1}) and its
implementation (\S\ref{theory2}).

\subsection{Model Description}
\label{theory1}
This model assumes the PWN is surrounded by a thin shell of material
with radius $R_{\rm pwn}$, mass $M_{\rm sw,pwn}$ and expansion
velocity $v_{\rm pwn}$ (e.g. \citealt{ostriker71, gelfand07}).  If
$v_{\rm pwn}$ is larger than the velocity of the material surrounding
this PWN, $v_{\rm ej}(R_{\rm pwn})$, we assume that $M_{\rm sw,pwn}$
increases by an amount $\Delta M_{\rm sw,pwn} = 4\pi R_{\rm pwn}^2
\Delta R_{\rm pwn} \rho_{\rm ej}(R_{\rm pwn})$, where $\rho_{\rm
ej}(R_{\rm pwn})$ is the density of material surrounding the PWN.  The
difference in pressure between the PWN interior to this mass shell,
$P_{\rm pwn}$, and surrounding SNR, $P_{\rm snr}(R_{\rm pwn})$,
applies a force, $F_{\Delta P}$, equal to \citep{gelfand07}:
\begin{equation}
\label{fdelp}
F_{\Delta P} = 4\pi R_{\rm pwn}^2 [ P_{\rm pwn} - P_{\rm snr}(R_{\rm pwn})].
\end{equation}
The resultant change in momentum of this shell of material is simply:
\begin{equation}
\frac{d}{dt}(M_{\rm sw,pwn}v_{\rm pwn}) = F_{\Delta P},
\end{equation}
and we use this equation to determine the dynamical evolution
(e.g. $R_{\rm pwn}(t)$ and $v_{\rm pwn}(t)$) of the PWN.  This
approach requires modeling the SNR's density, velocity, and pressure
profile with time and the pressure inside the PWN, $P_{\rm pwn}$.  We
evolve the properties of the SNR using the results of previously
developed analytic models described in Appendix \ref{snrtheory}.
Since no such descriptions exist for a SNR in the radiative phase of
its evolution, our model only predicts the evolution of a PWN inside a
SNR during its free-expansion and Sedov-Taylor evolutionary phases.

The evolution of $P_{\rm pwn}$ depends on the injection rate of energy
into the PWN by the central neutron star, the content of the pulsar
wind, and the evolution of the particle and magnetic components of the
PWN.  We assume that all of the spin-down luminosity $\dot{E}$ of the
neutron star is injected into the PWN (e.g. Equation 5 in
\citealt{gaensler06}):
\begin{equation}
\label{edoteqn}
\dot{E}(t) = \dot{E}_0 \left(1+\frac{t}{\tau_{\rm sd}}
  \right)^{-\frac{p+1}{p-1}},
\end{equation}
where $p$ is the the braking index\footnote{The braking index $p$ is
defined as $\dot{\Omega}_{\rm psr} = -k\Omega_{\rm psr}^p$ where
$\Omega_{\rm psr}$ is the spin frequency of the neutron star
($\Omega_{\rm psr} \equiv 2\pi/P$ where $P$ is the rotational period
of the neutron star; e.g. \citealt{shapiro86}).}, $\dot{E}_0$ is the
initial spin-down luminosity of the neutron star, and $\tau_{\rm sd}$
is the spin-down timescale.  As mentioned in \S\ref{intro}, it is
believed that neutron star will eventually be stripped of the PWN.
When this occurs, it no longer injects energy into the ``relic'' PWN
but forms a new PWN inside the SNR \citep{vdswaluw04}.  Based on the
simulations of \citet{vdswaluw04}, we assume the neutron star leaves
the PWN only after the PWN has collided with the SNR reverse shock,
when the distance the neutron star has traveled since the supernova
explosion, $r_{\rm psr}$, satisfies $r_{\rm psr}>R_{\rm psr}$.

The energy of the pulsar wind is distributed between the electrons and
positrons ($\dot{E}_{\rm inj,e}$), ions ($\dot{E}_{\rm inj,i}$), and
magnetic fields ($\dot{E}_{\rm inj,B}$) that comprise this outflow,
such that:
\begin{equation}
\label{edotcompeqn}
\dot{E} = \dot{E}_{\rm inj,e} + \dot{E}_{\rm inj,i} + \dot{E}_{\rm inj,B}.
\end{equation}
To parameterize the content of the pulsar wind, we define the
following variables:
\begin{eqnarray}
\label{etae}
\eta_e(t) & \equiv & \frac{\dot{E}_{\rm inj,e}(t)}{\dot{E}(t)} \\
\label{etai}
\eta_i(t) & \equiv & \frac{\dot{E}_{\rm inj,i}(t)}{\dot{E}(t)} \\
\label{etab}
\eta_{\rm B}(t) & \equiv & \frac{\dot{E}_{\rm inj,B}(t)}{\dot{E}(t)},
\end{eqnarray}
where $\eta_e + \eta_i + \eta_{\rm B} \equiv 1$.  It is important to
emphasize the $\eta_{\rm B}$ is {\it not} equivalent to the
magnetization parameter, $\sigma$, defined as the ratio of magnetic to
particle energy \citep{kennel84}.  The pulsar wind is injected into
the PWN at structure called the ``termination shock'', located where
the ram pressure of the unshocked wind is equal to $P_{\rm pwn}$
(e.g. \citealt{goldreich69}).  This occurs at a distance from the
neutron star, $r_{\rm ts}$, equal to (e.g. \citealt{slane04,
gaensler06}):
\begin{equation}
\label{radts}
r_{\rm ts} = \sqrt{\frac{\dot{E}}{4\pi\xi c P_{\rm pwn}}}
\end{equation}
where $\xi$ is the filling factor of the pulsar wind (an isotropic
wind has $\xi=1$).  In this model, we only concern ourselves with the
content of the pulsar wind just downstream of the termination shock
($r>r_{\rm ts}$), which is likely very different than the content of
the pulsar wind near the neutron star (e.g. \citealt{arons07}).

Since the energy of the pulsar wind is divided among electrons and
positrons, ions, and magnetic fields, the same must be true for the
energy inside the PWN, $E_{\rm pwn}$:
\begin{equation}
\label{epwneqn}
E_{\rm pwn} = E_{\rm pwn,e} + E_{\rm pwn,i} + E_{\rm pwn,B}
\end{equation}
where $E_{\rm pwn,e}$ is the kinetic energy of electron and positrons
in the PWN, $E_{\rm pwn,i}$ is the kinetic energy of ions in the PWN,
and $E_{\rm pwn,B}$ is the energy stored in the magnetic field of the
PWN.  Each component contributes separately to the total pressure
inside the PWN, such that $P_{\rm pwn}$ is:
\begin{equation}
\label{ppwn}
P_{\rm pwn} = P_{\rm pwn,e} + P_{\rm pwn,i} + P_{\rm pwn,B},
\end{equation}
where $P_{\rm pwn,e}$ is the pressure associated with electrons and
positrons in the PWN, $P_{\rm pwn,i}$ is the pressure associated with
ions in the PWN, and $P_{\rm pwn,B}$ is the pressure associated with
the PWN's magnetic field.  The energy and associated pressure of these
components evolve differently from each other, as explained below.

In this model, we assume the magnetic field inside the PWN is uniform
and isotropic.  As a result, $E_{\rm pwn,B}$ is:
\begin{equation}
\label{epwnb}
E_{\rm pwn,B} = \frac{B_{\rm pwn}^2}{8\pi}V_{\rm pwn}
\end{equation}
where $B_{\rm pwn}$ is the strength of the PWN's magnetic field, and the
pressure associated with this magnetic field $P_{\rm pwn,B}$ is:
\begin{equation}
\label{ppwnb}
P_{\rm pwn,B} = \frac{B_{\rm pwn}^2}{8\pi}.
\end{equation}
We also assume that the magnetic flux of the PWN is conserved as
$R_{\rm pwn}$ changes.  As a result, $B_{\rm pwn} \propto R_{\rm
pwn}^{-2}$ neglecting any input from the neutron star.  In this case, 
$E_{\rm pwn,B} \propto R_{\rm pwn}^{-1}$ from Equation \ref{epwnb} and
$P_{\rm pwn,B} \propto R_{\rm pwn}^{-4}$ from Equation \ref{ppwnb}.

We additionally assume that the electrons, positrons, and ions inside
the PWN are relativistic.  Since the density inside the PWN, it is
same to assume that these particles are essentially collisionless and
therefore behave as an ideal gas with adiabatic index $\gamma=4/3$.
Therefore, the pressure associated with particles inside the PWN
$P_{\rm pwn,p}$ is:
\begin{equation}
\label{ppwnpar}
P_{\rm pwn,p} = \frac{E_{\rm pwn,p}}{3 V_{\rm pwn}},
\end{equation}
where $E_{\rm pwn,p}$ is the total particle energy of the PWN ($E_{\rm
pwn,p} \equiv E_{\rm pwn,e} + E_{\rm pwn,i}$) and $V_{\rm pwn}$ is the
volume of the PWN ($V_{\rm pwn} \equiv \frac{4}{3}\pi R_{\rm pwn}^3$).
The evolution of $E_{\rm pwn,p}$ and $P_{\rm pwn,p}$ depends on
adiabatic and radiative losses suffered by the particles inside the
nebula.  Since we are assuming that all the particles inside the PWN
are relativistic (all three populations has $\gamma=4/3$), the
adiabatic evolution of $E_{\rm pwn,p}$ is independent of the
distribution of particle energy between these species as well as their
energy spectrum.  If the evolution of the PWN is purely adiabatic,
then $P_{\rm pwn,p}V_{\rm pwn}^{4/3}$ is constant, implying that
$E_{\rm pwn,p} \propto R_{\rm pwn}^{-1}$ and $P_{\rm pwn,p} \propto
R_{\rm pwn}^{-4}$.

The total radiative losses of electrons/positrons and ions in the PWN
depends greatly on the energy distribution and is sensitive to their
individual energy spectra.  Their energy spectrum is highly sensitive
to the injection spectrum of particles into the PWN.  Observations of
several PWNe (e.g.  3C58; \citealt{slane08}, PWN G0.9+0.1;
\citealt{venter06}) suggest a broken power-law injection spectrum for
electrons and positrons, and recent theoretical work
(e.g. \citealt{spitkovsky08}) argues the injection spectrum of these
particles is a Maxwellian with a non-thermal tail.  For simplicity, we
assume a simple power law injection spectrum for the electrons,
positrons, and ions, but our formalism can easily accommodate a more
complex input spectrum.

Assuming that the injection spectrum of electrons and positrons obeys
a single power-law, we have:
\begin{equation}
\label{elecspecdist}
n_{e} = n_{0,e} \left( \frac{E}{E_0} \right)^{-\gamma_e} {\rm
  electrons~s^{-1}~keV^{-1}},
\end{equation}
where $n_{e}\Delta E \Delta t$ is the number of electrons and
positrons with energy between $E$ and $E+\Delta E$ injected into the
PWN in time $\Delta t$.  For consistency with Equation
\ref{edotcompeqn}, we require that:
\begin{equation}
\label{edotelec}
\dot{E}_{\rm inj,e} \equiv \int_{E_{e, \rm min}}^{E_{e, \rm max}}
n_{e} E dE,
\end{equation}
where $E_{e, \rm min}$ and $E_{e, \rm max}$ are, respectively, the
minimum and maximum energy of electrons and positrons injected into
the PWN. Therefore, the rate of electrons and ions injected into the
PWN, $\dot{N}_{\rm inj,e}$, is:
\begin{equation}
\label{ndotelec}
\dot{N}_{\rm inj,e} = \left\{\begin{array}{ll}
 \left(\frac{2-\gamma_e}{1-\gamma_e}\right) \frac{E_{e, \rm
 max}^{1-\gamma_e} - E_{e, \rm min}^{1-\gamma_e}}{E_{e, \rm
 max}^{2-\gamma_e} - E_{e, \rm min}^{2-\gamma_e}} \dot{E}_{\rm inj,e}
 & \gamma_e \neq 1,2 \\ 
\frac{E_{e, \rm min}^{-1} - E_{e,\rm max}^{-1}}{\ln(E_{e, \rm
 max}/E_{e, \rm min})} \dot{E}_{\rm inj,e} &  \gamma_e=2 \\
\frac{\ln(E_{e, \rm max}/E_{e, \rm min})}{E_{e, \rm max}-E_{e, \rm min}}
\dot{E}_{\rm inj,e} & \gamma_e=1
\end{array}
   \right. .
\end{equation}
Similarly, we assume the injection spectrum of ions is a single power
law:
\begin{equation}
\label{ionspecdist}
n_{i} = n_{0,i} \left( \frac{E}{E_0} \right)^{-\gamma_i},
\end{equation}
where $n_{i} \Delta E \Delta t$ is the number of ions with energy
between $E$ and $E+\Delta E$ injected into the PWN in time $\Delta t$.
Again, for consistency with Equation \ref{edotcompeqn} we require
that:
\begin{equation}
\label{edotion}
\dot{E}_{\rm inj,i} \equiv \int_{E_{i, \rm min}}^{E_{i, \rm max}} n_{i} EdE,
\end{equation}
where $E_{i, \rm min}$ and $E_{i, \rm max}$ are the minimum and
maximum energy of ions injected into the PWN.  This results in the
injection rate of ions into the PWN, $\dot{N}_{\rm inj,i}$, having the
same functional form as Equation \ref{ndotelec}.

This model assumes the dominant radiative processes are synchrotron
emission and inverse Compton scattering off background photons.  For a
particle with energy $E$, mass $m$, and charge $q$, the rate at which
it loses energy due to synchrotron emission, $P_{\rm synch}$, is
(Equation 3.32 in \citealt{pacholczyk70}):
\begin{equation}
\label{psyncheqn}
P_{\rm synch} = \frac{2q^4}{3m^4c^7}B_{\rm pwn}^2 \sin^2 \theta E^2,
\end{equation}
where $\theta$ is the angle between the particle's velocity and the
magnetic field.  We assume that the velocities of particles in the PWN
are randomly oriented with respect to the magnetic field, such that
$\sin^2\theta=2/3$ \citep{rybicki79}.  When the PWN is small,
synchrotron self-absorption is important (e.g. \citealt{reynolds84}).
To determine its effect on the dynamics and emission of the PWN, we
first assume that a particle emits all of its synchrotron radiation at
a frequency $\nu=\nu_{\rm crit}$, where $\nu_{\rm crit}$ is equal to
(Equation 3.28 in \citealt{pacholczyk70}):
\begin{equation}
\label{nucrit}
\nu_{\rm crit} = \frac{3q}{4\pi m^3c^5}B_{\rm pwn} E^2 \sin\theta,
\end{equation}
where $\sin\theta=\sqrt{2/3}$.  The optical depth of the PWN to this
emission is (\S3.4 in \citealt{pacholczyk70}):
\begin{equation}
\label{taussa}
\tau(\nu_{\rm crit}) = 4.2\times10^{7} \frac{(B_{\rm pwn} \sin
\theta)^{3/2}}{\nu_{\rm crit}^{5/2}} \rho_p R_{\rm pwn} K(1),
\end{equation}
where $\sin\theta=\sqrt{2/3}$ as before, $\rho_p$ is the density of
particles inside the PWN, and $K(1)=1.1$ \citep{pacholczyk70}.  We
assume that particles with energy such that $\tau(\nu_{\rm crit}) >
1$ do not lose energy through synchrotron radiation -- an approach
similar to that of \citet{reynolds84}.

The power lost by particles due to inverse Compton scattering off
background photons, $P_{\rm IC}$, is (Equation 5.57 in
\citealt{pacholczyk70}):
\begin{equation}
\label{piceqn}
P_{\rm IC} = \frac{32\pi c q^4}{9(mc^2)^4}u_{\rm rad} E^2 f(E),
\end{equation}
where $u_{\rm rad}$ is the energy density of the background photon
field, and $f(E)$ is the inverse Compton scattering cross-section
relative to the Thomson cross-section.  $f(E)$ is sensitive to both
the particle energy and the spectrum of the background photon field.
The background photons are believed to be dominated by the Cosmic
Microwave Background (CMB; $u_{\rm rad} \approx 4.17\times10^{-13}$
ergs cm$^{-3}$), and we calculate $f(E)$ for this field using the
procedure described in \S2.3 of \citet{volpi08} (Figure
\ref{powicratfig}).  It is possible that other photon fields,
e.g. starlight, not considered here are important for a given PWN --
and a simple adjustment to the formalism presented here can account
for these as well.

\subsection{Model Implementation}
\label{theory2}
To determine the evolution of a PWN inside a SNR, we use the equations
presented in \S\ref{theory1} to determine how the relevant properties
of a PWN at a time $t$ evolve to a time $t+\Delta t$.  The procedure
used to determine the initial conditions of the PWN inside a SNR is
described in Appendix \ref{initialcond}, and the free parameters of
this model are listed in Table \ref{modelinput}.  We calculate the
properties of the PWN at time $t+\Delta t$ using the following
procedure:
\begin{itemize}
\item {\bf Step 1:} Calculate the new radius of the PWN
  \begin{equation}
    R_{\rm pwn}(t+\Delta t) = R_{\rm pwn}(t) + v_{\rm pwn}(t) \times
    \Delta t,
  \end{equation}
  and the distance the neutron star has traveled inside the SNR since
  birth:
  \begin{equation}
    r_{\rm psr}(t + \Delta t) = v_{\rm psr}(t + \Delta
    t)\times(t+\Delta t),
  \end{equation}
  where $v_{\rm psr}$ is the space velocity of the neutron star.

\item {\bf Step 2:} Adjust the electron/positron, ion, and magnetic
  field energy of the PWN for the adiabatic work done by the PWN on
  its surroundings and, if the neutron star is still inside the PWN,
  the energy it injected between times $t$ and $t+\Delta t$:
  \begin{eqnarray}
    E_{\rm pwn,e}(t+\Delta t) & = & \frac{R_{\rm pwn}(t)}{R_{\rm pwn}(t +
      \Delta t)} E_{\rm pwn,e}(t)+ \eta_e \int_{t}^{t+\Delta t} \dot{E} dt \\
    E_{\rm pwn,i}(t+\Delta t) & = & \frac{R_{\rm pwn}(t)}{R_{\rm pwn}(t +
      \Delta t)} E_{\rm pwn,i}(t) + \eta_i \int_{t}^{t+\Delta t} \dot{E} dt \\
    E_{\rm pwn,B}(t+\Delta t) & = & \frac{R_{\rm pwn}(t)}{R_{\rm pwn}(t +
      \Delta t)} E_{\rm pwn,B}(t) + \eta_{\rm B} \int_{t}^{t+\Delta t}
      \dot{E} dt 
  \end{eqnarray}
  where $\dot{E}$ is defined in Equation \ref{edoteqn} and $\int
  \dot{E} dt$ is solved analytically.  We then use $E_{\rm
  pwn,B}(t+\Delta t)$ to calculate $B_{\rm pwn}(t+\Delta t)$ using
  Equation \ref{epwnb} and $P_{\rm pwn,B}(t+\Delta t)$ using Equation
  \ref{ppwnb}.
  
\item {\bf Step 3}: Calculate the energy spectrum of electrons,
  positrons, and ions inside the PWN at time $t+\Delta t$ taking into
  account adiabatic, synchrotron, and inverse Compton losses.  A
  particle with energy $E$ at time $t$ has an energy at time $t+\Delta
  t$ equal to:
  \begin{equation}
    E(t+\Delta t) = \frac{R_{\rm pwn}(t)}{R_{\rm pwn}(t + \Delta t)}E -
    [P_{\rm synch}(t) + P_{\rm IC}(t)]\Delta t
  \end{equation}
  where $P_{\rm synch}(t)$ (Equation \ref{psyncheqn}) is calculated
  using $B_{\rm pwn}(t)$ and $P_{\rm IC}(t)$ is defined in Equation
  \ref{piceqn}, with the restriction that $E(t+\Delta t)\geq 0$.  We
  then subtract from the value of $E_{\rm pwn,e}(t+\Delta t)$ and
  $E_{\rm pwn,i}(t+\Delta t)$ calculated in Step 3 the total
  synchrotron and inverse Compton losses of these particles. Even
  though the radiative losses of ions in the PWN is significantly
  smaller than that of electrons and positrons (both $P_{\rm synch}$
  and $P_{\rm IC}$ are $\propto m^{-4}$ ), for completeness they are
  calculated.  We then determine $P_{\rm pwn,p}(t+\Delta t)$ using
  Equation \ref{ppwnpar}.

\item {\bf Step 4}: Calculate the properties of the surrounding SNR
  [$\rho_{\rm ej}(R_{\rm pwn},t+\Delta t)$, $v_{\rm ej}(R_{\rm pwn},t
  + \Delta t)$, and $P_{\rm snr}(R_{\rm snr},t+\Delta t)$] using the
  prescription described in Appendix \ref{snrtheory}.  If $v_{\rm ej}(R_{\rm
  pwn},t + \Delta t) < v_{\rm pwn}(t)$ then:
  \begin{equation}
    M_{\rm sw,pwn}(t+\Delta t) = M_{\rm sw,pwn}(t) + \frac{4}{3} \pi
     \left[R_{\rm pwn}(t+\Delta t)^3 - R_{\rm pwn}(t)^3 \right]
     \rho_{\rm ej}(R_{\rm pwn},t+\Delta t).
  \end{equation}
  Otherwise, $M_{\rm sw,pwn}(t+\Delta t)=M_{\rm sw,pwn}(t)$.

\item {\bf Step 5}: Using the properties of the PWN and SNR determined
  above, calculate $F_{\Delta P}(t)$ (Equation \ref{fdelp}).  The new
  velocity of the mass shell surrounding the PWN, $v_{\rm
  pwn}(t+\Delta t)$ is :
  \begin{equation}
    v_{\rm pwn}(t+\Delta t) = \frac{1}{M_{\rm sw,pwn}(t+\Delta t)}
    \times [M_{\rm sw,pwn}(t)v_{\rm pwn}(t) + \Delta M_{\rm
    sw,pwn}v_{\rm ej}(R_{\rm pwn},t) + F_{\rm pwn}(t) \times \Delta
    t],
  \end{equation}
  from conservation of momentum arguments, where $\Delta M_{\rm
  sw,pwn} \equiv M_{\rm sw,pwn}(t+\Delta t) - M_{\rm sw,pwn}(t)$.
\end{itemize}

\section{Model Performance} 
\label{performance}
In this Section, we present the evolution of a PWN inside a SNR
predicted by this model for the combination of the input variables
listed in Table \ref{trialmodel}.  It is important to emphasize that
these results are specific to this set of parameters, and the
different PWNe may have very different values for some or all of these
quantities.  The properties of the progenitor supernova ($E_{\rm sn}$
and $M_{\rm ej}$), central neutron star ($\dot{E}_0$, $\tau_{\rm sd}$,
and $p$), and surrounding ISM ($n_{\rm ism}$) are the same as Model A
in \citet{blondin01} -- chosen by these authors to reproduce a PWN
similar to that of the Crab Nebula.  We also assume a purely
electron/positron pulsar wind (i.e. $\eta_i\equiv0$), and that the
properties of the pulsar wind ($\eta_B$, $\eta_e$, $E_{\rm e,min}$,
$E_{\rm e,max}$, $\gamma_e$) remain constant with time.  We also use a
pulsar velocity of 120 km s$^{-1}$, the most recent measurement of the
transverse velocity of the Crab pulsar \citep{kaplan08}.  As shown in
Figure \ref{radpwnfig}, for this particular set of input parameters
our model predicts four evolutionary stages:
\begin{itemize}
\item {\bf Initial expansion:} This phase ends $t_{\rm col}\sim4500$
years after the supernova explosion when the PWN collides with the
reverse shock.
\item {\bf Reverse Shock Collision and First Compression:} This phase
ends $\sim20000$ years after the supernova explosion.  During this
phase, $\sim17000$ years after the supernova explosion, we expect the
PWN to be stripped of its neutron star.
\item {\bf Re-expansion:} This phase ends $\sim56000$ years after the
supernova.  During this phase, the neutron star re-enters the
``relic'' PWN created during the first compression $\sim30000$ years
after the supernova explosion.
\item {\bf Second Compression:} This phase continues until the SNR
enters the radiative phase of its evolution.  During this contraction,
the PWN is stripped of its neutron star $\sim70000$ years after the
supernova explosion.
\end{itemize}
\noindent The dynamical and radiative evolution of the PWN in these
four phases is described in detail below.

The initial expansion of the PWN is the result of $P_{\rm pwn}\gg
P_{\rm snr}(R_{\rm pwn})$ (Figure \ref{pressfig}).  This is the case
because $P_{\rm snr}(R_{\rm pwn})\approx0$ when $R_{\rm pwn}<R_{\rm
rs}$, i.e. adiabatic expansion of the SNR makes the ejecta interior of
the reverse shock cold (Appendix \ref{snrtheory}).  During this time,
the PWN is expanding faster than the surrounding ejecta (Figure
\ref{velfig}), and $M_{\rm sw,pwn}$ increases (Figure \ref{mswfig}).
For most of this expansion, the energy input for the neutron star
($\dot{E}$) is larger than the adiabatic losses resulting from the
PWN's expansion and the radiative losses suffered by electrons inside
the PWN (Figure \ref{edotfig}), causing the internal energy of the PWN
to rise (Figure \ref{epwnfig}).  At the very earliest times ($t<100$
years), synchrotron losses approach the energy input of the pulsar due
to the PWN's very strong magnetic field (Figure \ref{bpwnfig}).  These
synchrotron losses cause the magnetization parameter of the PWN,
$\sigma$, to temporarily increase by more than an order of magnitude
(Figure \ref{sigmafig}).  For $t\la1000$ years, synchrotron losses
actually dominate over adiabatic and inverse Compton losses (Figure
\ref{edotfig}), causing the PWN to expand slower than predicted by
Equation \ref{rpwninit}.  This difference is small at early times, so
Equation \ref{rpwninit} does provides a reasonably accurate initial
condition.  The expansion of the PWN causes $B_{\rm pwn}$ to decrease
rapidly (Figure \ref{bpwnfig}), resulting in a rapid decrease of the
PWN's synchrotron luminosity.  As a result, adiabatic losses dominate
from $\sim1000$ years after the supernova explosion until the PWN
collides with the reverse shock. These adiabatic losses cause both
$B_{\rm pwn}$ and $E_{\rm pwn,B}$ to decrease for $t>1000$ years,
while $E_{\rm pwn,p}$ increases due to the continued inject of
particles into the PWN by the pulsar.  It is important to note that
$B_{\rm pwn}$ does {\it not} follow $B_{\rm pwn}\propto
1/(1+(t/\tau_{\rm sd})^\alpha)$, as assumed in other work
(e.g. \citealt{venter06}).  We find that $B_{\rm pwn}\propto t^{-1.7}$
during the initial expansion, similar to the $t^{-1.3}-t^{-2}$
behavior derived by \citet{reynolds84}.

This behavior ends when the PWN collides with the SNR reverse shock.
This collision shocks the swept-up ejecta surrounding the PWN, but not
the PWN itself because the sound speed inside the PWN ($\sim
c/\sqrt{3}$) is significantly higher than the velocity of the reverse
shock.  The collision with the reverse shock marks the end of PWN's
rapid expansion (Figure \ref{radpwnfig}) because the PWN is no longer
in an essentially pressureless environment.  At the time of this
collision, $P_{\rm pwn}\ll P_{\rm snr}(R_{\rm pwn})$ (Figure
\ref{pressfig}), causing $v_{\rm pwn}$ to decrease significantly
(Figure \ref{velfig}).  However, for the first $\sim100$ years after
this collision, $v_{\rm pwn}>v_{\rm ej}(R_{\rm pwn})$ (Figure
\ref{velfig}), resulting in $M_{\rm sw,pwn}$ continuing to rise,
increasing from $\sim1 M_{\odot}$ to $\sim3 M_{\odot}$ (Figure
\ref{mswfig}) due to the high density behind the reverse shock.  The
sharp decrease in $v_{\rm pwn}$ also leads to a sharp decrease in
adiabatic losses (Figure \ref{edotfig}).  Because the pulsar continues
to inject energy into the PWN, both $E_{\rm pwn,p}$ and $E_{\rm
pwn,B}$ initially rise rapidly after the PWN/reverse shock collision
(Figure \ref{epwnfig}), though $P_{\rm pwn}$ is still $P_{\rm pwn} \ll
P_{\rm snr}(R_{\rm pwn})$.  As a result, the high-pressure ejecta
downstream of the reverse shock will compress the PWN beginning
$\sim6500$ years after the supernova explosion.

This compression begins a series of contractions and re-expansions
that continue until the SNR enters the radiative phase of its
evolution (Figure \ref{radpwnfig}).  Similar behavior was observed in
previous work (e.g. \citealt{blondin01, bucciantini04, vdswaluw01b,
vdswaluw04}).  The adiabatic compression of the PWN causes a sharp
rise in the internal pressure (Figure \ref{pressfig}) and the particle
and magnetic energy (Figure \ref{epwnfig}) of the PWN.  The
compression of the PWN also causes a rapid rise in $B_{\rm pwn}$
(Figure \ref{bpwnfig}), which in turn results in a rapid rise in the
PWN's synchrotron luminosity (Figure \ref{edotfig}) -- as previously
noted by \citet{reynolds84}.  Eventually, the synchrotron losses are
larger than the work done by the surrounding SNR on the PWN and the
rate particle energy is injected into the PWN by the neutron star,
causing $E_{\rm pwn,p}$ to decrease (This would have occurred during
the second compression if we evolved the PWN further in time).  This
decrease in $E_{\rm pwn,p}$ leads to a decrease in the synchrotron
luminosity of the PWN (Figure \ref{edotfig}), even though the magnetic
field inside the PWN is still increasing (Figure \ref{bpwnfig}).
Since synchrotron losses do not diminish $E_{\rm pwn,B}$, $\sigma$
increases significantly during the compression phases of the PWN
(Figure \ref{sigmafig}).

Due to the space velocity of the neutron star, we expect that the PWN
will be stripped of its pulsar during both compressions.  It is
important to note that this result is strongly dependent on the space
velocity of the pulsar.  For a lower space velocity, it is possible
that the PWN will not be stripped of its pulsar until the pulsar
leaves the SNR, while for a higher velocity the PWN will not overtake
the pulsar during its re-expansion.  The departure of the pulsar
causes a rapid increase of $\sigma$ (Figure \ref{sigmafig}) and a
rapid decrease in the synchrotron luminosity of the PWN since it
removes the only source of high energy particles that dominate this
emission (Figure \ref{edotfig}).  The latter does not occur during the
second compression due to the rapid rise in $B_{\rm pwn}$ when the
pulsar exits the PWN. It is expected that the pulsar creates a new PWN
inside the SNR after it leaves its original PWN
(e.g. \citealt{vdswaluw04}) whose properties we do not calculate.

Eventually during the compression stage, the pressure inside the PWN
will exceed that of the surrounding SNR, causing the PWN to re-expand.
During the first contraction, the pressure inside the PWN exceeds that
of the surrounding SNR $\sim5000$ years after the compression begins
(Figure \ref{pressfig}).  At this time, the PWN is contracting at a
very high velocity ($>1000$ km s$^{-1}$; Figure \ref{velfig}) and, due
to the considerable momentum of the swept-up material surrounding the
PWN, it takes $\sim8000$ years for PWN to re-expand (Figure
\ref{radpwnfig}).  Therefore, $P_{\rm pwn}\gg P_{\rm snr}(R_{\rm
pwn})$ when re-expansion begins (Figure \ref{pressfig}).  During this
phase, the PWN sweeps up additional material.  Due to the low density
inside the SNR at this time, the mass shell only accumulates $\sim0.5
M_{\odot}$ of additional material.  Initially during the re-expansion,
$E_{\rm pwn,p}$, $E_{\rm pwn,B}$ (Figure \ref{edotfig}), and $B_{\rm
pwn}$ (Figure \ref{bpwnfig}) rapidly decrease due to adiabatic losses.
The decrease in $E_{\rm pwn}$ and increase in $R_{\rm pwn}$ result in
a rapid decrease in $P_{\rm pwn}$ (Figure \ref{pressfig}), and the
decrease in both $B_{\rm pwn}$ and $E_{\rm pwn,p}$ result in a rapid
decrease of the PWN's synchrotron luminosity (Figure \ref{edotfig}).
During the re-expansion, $\sigma$ initially continues to increase
because the synchrotron and adiabatic losses suffered by particles in
the PWN are greater than the adiabatic losses of the magnetic field
(Figure \ref{sigmafig}).

During the re-expansion, $v_{\rm pwn}$ reaches $\sim300$\ km s$^{-1}$,
higher than the space velocity of the neutron star.  As a result, the
pulsar will re-enter the ``relic'' PWN.  At the time of re-entry
$\dot{E}$ is actually higher than the synchrotron, adiabatic, and
inverse Compton losses of the PWN (Figure \ref{edotfig}), causing
$E_{\rm pwn}$ to increase (Figure \ref{epwnfig}).  Since most of the
energy deposited by the pulsar in the PWN is in the form of particles
($\eta_e \gg \eta_{\rm B}$), only $E_{\rm pwn,p}$ increases as a
result of the re-entry of the pulsar -- $\dot{E}_{\rm inj,B}$ is
significantly less than the adiabatic losses suffered by the magnetic
field (Figure \ref{epwnfig}).  As a result, $\sigma$ begins to
decrease, though $\sigma\gg\eta_B$ at these times.  The increase in
$E_{\rm pwn,p}$ also leads to a sharp rise in both the synchrotron and
inverse Compton luminosity of the PWN (Figure \ref{edotfig}).  This
rise in energy is {\it not} accompanied by a rise in pressure (Figure
\ref{pressfig}) because the volume of the PWN is increasing faster
than the particle energy.  As a result, $\sim30000$ years after the
supernova explosion $P_{\rm pwn}<P_{\rm snr}(R_{\rm pwn})$, and the
PWN will be compressed for a second time.  As shown in Figure
\ref{radpwnfig}, this second compression is expected to continue until
the SNR enters the radiative phase of its evolution, and the evolution
of the PWN during this phase is described above.

The complicated dynamical evolution of a PWN described above causes
the energy spectrum of electrons and positrons in the PWN, and the
photons they radiate, to change considerably with time.  As mentioned
earlier, synchrotron losses play a very important role during the
initial expansion of the PWN.  For $t \ll \tau_{\rm sd}$, the magnetic
field of the PWN is so strong that the only high energy ($E\ga1$~TeV)
electrons and positrons in the PWN are those which were injected very
recently (Figure \ref{elecspec}).  For $\tau_{\rm sd} < t < t_{\rm
col}$, the electron and positron spectrum of the PWN is dominated by
previously injected particles at all energies, and the sharp
high-energy cutoff observed at early times is replaced with a more
gradual turnover whose sharpness increases with time (Figure
\ref{elecspec}).  This is due to the increase peak energy of the
electron and positron spectrum (as first noted by
\citealt{reynolds84}) resulting from the decrease in $B_{\rm pwn}$.

Correspondingly, the spectrum of photons radiated by the PWN during
its initial expansion also evolves with time (Figure \ref{photspec}).
As explained in \S\ref{theory}, the only emission mechanisms we
consider are synchrotron radiation and inverse Compton scattering off
the CMB, and the photon spectrum is calculated using the procedure
described in Section 2 of \citet{volpi08}.  At very early times
($t<50$ years), the PWN is extremely luminous at GeV energies due to
the large number of high energy electrons and positrons recently
injected by the neutron star (Figure \ref{photlumin}).  For
$t<\tau_{\rm sd}$, the PWN's synchrotron emission peaks at photon
energies $\ga 100$ keV (Figure \ref{photspec}), and the PWN's
luminosity is highest in the hard X-ray regime (Figure
\ref{photlumin}).  Since recently injected particles are energetically
important, both the synchrotron and inverse Compton spectrum have two
peaks -- a lower energy peak resulting from previously injected
particles and a higher energy peak resulting from recently injected
ones.  For $\tau_{\rm sd}<t<t_{\rm col}$, the synchrotron and inverse
Compton spectrum have a single peak since recently injected particles
no longer dominate.  During the initial expansion, the energy peak of
the synchrotron spectrum decreases from $\sim1$ MeV at $t\ll\tau_{\rm
sd}$ to $\sim1-10$ keV at $t\sim t_{\rm col}$, causing a rapid
decrease in the hard X-ray luminosity of the PWN.  Slower decreases
are predicted for the radio -- soft X-ray luminosity (Figure
\ref{photlumin}) due to the gradual decline in the synchrotron
luminosity of the PWN (Figure \ref{edotfig}) resulting from the
decreasing value of $B_{\rm pwn}$ (Figure \ref{bpwnfig}).  Conversely,
the energy peak of inverse Compton emission increases from $\sim1$ TeV
at $t\sim\tau_{\rm sd}$ to $\sim50$ TeV at $t\sim t_{\rm col}$ due to
the increase in the break energy of the electron spectrum discussed
above, leading to a rapid rise in the GeV and TeV $\gamma$-ray
luminosity of the PWN (Figure \ref{photlumin}).

The energy spectrum of electrons and positrons inside the PWN changes
significantly during the first contraction (Figure \ref{elecspec}).
Due to the strong magnetic field inside the PWN during this phase
(Figure \ref{bpwnfig}), the synchrotron lifetime of the highest energy
(E$\ga$10 TeV) electrons and positrons in the PWN becomes
significantly less than the age of the PWN.  This results in a sharp
cut-off in the electron and positron energy spectrum where the
synchrotron lifetime of electrons and positrons is the age of the PWN,
above which recently injected particles dominate.  The strengthening
$B_{\rm pwn}$ results in this energy decreasing with time.  This
causes the peak photon energy of the synchrotron emission to decrease
significantly during this time, from $\sim 10$ keV when the PWN
collides with the reverse shock to $\sim100$ eV when the neutron star
exist the PWN.  The spectrum of the inverse Compton emission radiated
by the PWN also changes considerably during this time -- with the
energy peak decreasing from $\sim10$ TeV at the time of the
PWN/reverse shock collision to $\sim100$ GeV when the neutron star
leaves.  The luminosity of the PWN in the wavebands dominated by
synchrotron emission (radio -- soft X-rays; Figure \ref{photspec})
increase (Figure \ref{photlumin}) due to the strengthening magnetic
field.  The hard X-ray and GeV $\gamma$-ray luminosity of the PWN
increases due to the decreasing energy of the inverse Compton peak,
which causes a decrease in the TeV $\gamma$-ray luminosity of the PWN
(Figure \ref{photlumin}) mitigated by inverse Compton emission from
the highest energy recently injected particles.

The departure of the neutron star from the PWN during the first
contraction has a dramatic effect on the electron and positron energy
spectrum in the PWN.  When this occurs, the only high energy ($E>10$
TeV) electrons and positrons in the PWN are those recently injected by
the neutron star.  Therefore, the departure of the neutron star causes
this plateau of high-energy electrons and positrons to quickly
disappear due to synchrotron cooling (Figure \ref{elecspec}).  This
results in a sharp decrease in the PWN's X-ray and TeV luminosity
(Figure \ref{photlumin}).  It is possible that the new PWN will be
luminous at these energies, so the total X-ray and TeV luminosity of
the system might be appreciably higher than predicted here.  The
energy peak of the electron and positron spectrum also decreases from
$E\sim10$ TeV when the neutron star leaves to $E\sim10$ GeV at the
time of re-expansion (Figure \ref{elecspec}), with the energy spectrum
highly suppressed above the peak.  This causes the peak of the
synchrotron emission to decrease from the ultra-violet to mid-IR
wavelengths and the bulk of the PWN's synchrotron emission to move
from the optical to radio wavelengths, while the peak of the inverse
Compton emission to decrease from $\sim100$ GeV to $\sim10$ MeV
(Figure \ref{photspec}), causing the PWN to be more luminous at GeV
energies then TeV energies (Figure \ref{photlumin}).

As the PWN re-expands into the SNR but before it overtakes the pulsar,
the maximum energy of electrons and positrons in the PWN decreases
from $\sim10$ GeV to $\sim1$ GeV (Figure \ref{elecspec}).  This fact,
and the decrease in $B_{\rm pwn}$ (Figure \ref{bpwnfig}), causes the
peak frequency of the PWN's synchrotron spectrum to decrease to $\sim
1$~GHz and the peak of the inverse Compton radiation to decrease to
$\sim100$ keV.  As a result, when the PWN is re-expanding inside the
SNR but before it overtakes the neutron star, almost all of its
emission is in the radio, soft X-ray, and hard X-ray bands (Figure
\ref{photlumin}).  When the neutron star re-enters the PWN, it resumes
injecting high-energy electrons and positrons into the PWN --
resulting in an electron spectrum with two distinct components: a
lower energy population composed of electrons and particle injected at
earlier times and a higher energy population composed of recently
injected particles (Figure \ref{elecspec}).  During the PWN's
re-expansion, the peak energy of the ``old'' particles continues to
decrease due to adiabatic losses while the peak energy of the ``new''
particles decreases due to synchrotron losses (Figure \ref{elecspec}).
This dichotomy also extends to the photon spectrum (Figure
\ref{photspec}).  The radio emission from the PWN is dominated by
synchrotron radiation from the old particles, and the radio luminosity
of the PWN decreases (Figure \ref{photlumin}) due to their decreasing
energy.  At all other wavebands, the emission is dominated by recently
injected particles.  Initially, the synchrotron emission from these
electrons and positrons extends to $\sim100$ keV, but due to adiabatic
and synchrotron losses this decreases to $\sim10$ keV by the PWN
begins its second contraction (Figure \ref{photspec}).  As a result,
after an initial increase which marks the neutron star's re-entry,
both the soft X-ray and hard X-ray luminosity of the PWN will decrease
-- though the decline in the soft X-ray luminosity is significantly
slower than the hard X-ray luminosity (Figure \ref{photlumin}).  The
inverse Compton emission from the recently injected electrons and
positrons extends to TeV energies (Figure \ref{photspec}), and the
$\gamma$-ray luminosity of the PWN increases between the re-entry of
the pulsar and the second compression (Figure \ref{photlumin}).

During the second contraction of the PWN, the energy spectrum of
electrons and positrons inside the PWN maintains the two component
structure created during the PWN's re-expansion (Figure
\ref{elecspec}).  The peak energy of the ``old'' electron population
increases slightly from $\sim1$ GeV to $\sim5$ GeV due to the work
done on the PWN by the surrounding ejecta, but the energy peak of the
``new'' electron population decreases from $\sim100$ TeV to $\sim1$
TeV (Figure \ref{elecspec}) due to the increasing strength of the
PWN's magnetic field and the second departure of the pulsar from the
PWN $\sim15000$ years after the second contraction begins.  During the
second contraction, the photon spectrum of the PWN has four distinct
regimes -- from lowest to highest energy, they are synchrotron
emission from ``old'' electrons and positrons, the synchrotron
emission from ``new'' electrons and positrons, the inverse Compton
emission from ``old'' particles, and the inverse Compton emission from
``new'' particles (Figure \ref{photspec}).  During the second
compression, the energy peaks corresponding to emission from the old
particles increase -- the peak frequency of the synchrotron emission
rises from $\nu\sim10$~MHz to $\nu\sim1$~GHz, while the peak energy of
their inverse Compton emission rises from $\sim10$~keV to
$\sim$100~keV.  This results in an increase in the radio and hard
X-ray luminosity of the PWN (Figure \ref{photlumin}).  Conversely, the
energy peaks of the spectral features produced by the recently
injected particles decreases -- the synchrotron peak decreases from
$\sim10$ keV to $\sim1$ eV, while the inverse Compton peak decreases
from $\sim100$ TeV to $\sim10$ GeV, causing an increase in the optical
and mid-IR luminosity and a decrease in the $\gamma$-ray luminosity of
the PWN.

The observable properties discussed above are calculated using the
predictions of the model.  Unfortunately, for most PWNe the only
observational data available is a flux density measurement at two or
three radio frequencies (typically, 1.4, 4.8, and 8.5 GHz) and a
measurement of the X-ray spectrum between $E_{\gamma} \sim 0.5-10$ keV
-- though due to interstellar absorption, for several PWN it is only
possible to measure the spectrum between $E_{\gamma}=2-10$ keV
(e.g. G328.4+0.2; \citealt{gelfand07}).  This information is then used
to estimate the properties of the PWN, for example its energetics and
strength of its magnetic field.  The radio and unabsorbed X-ray
spectrum are usually fit to a power law, $L_{\nu} \propto
\nu^{\alpha}$, where $\alpha$ is the spectral index.  (The X-ray
spectrum is often fit using $N_{\nu} \propto \nu^{-\Gamma}$, where
$N_{\nu}$ is the number density of observed photons and $\Gamma$ is
the photon index, where $\alpha=1-\Gamma$.)  We derive the spectral
index of the PWN in the radio, $0.5-10$ keV, and $2-10$ keV bands
throughout its evolution, finding that the radio spectral index is
$\approx-0.3$ except for two periods where $\alpha \ll -1.6$ (Figure
\ref{alphafig}).  These two period are during the re-expansion of the
PWN when the frequency of the lower-energy synchrotron peak decreases
from $\nu\gg8.5$ GHz to $\nu<1.4$ GHz and during the second
contraction of the PWN when the low-energy synchrotron peak
transitions from $\nu<1.4$ GHz to $\nu>8.5$ GHz (Figure
\ref{photdensfig}).  The photon index $\Gamma$ measured between
0.5--10 keV and 2--10 keV is considerably more variable, fluctuating
between $\Gamma\sim1$ and $\Gamma\gg3$.  As a result, the change in
spectral index ($\Delta\alpha$) between the radio and the two X-ray
bands varies significantly with time (Figure \ref{delalphafig}), with
$\Delta\alpha$ often $\Delta\alpha>0.5$, the value expected from
standard synchrotron theory for a constant magnetic field
\citep{pacholczyk70}.

There are several different ways to estimate the internal energy and
strength of the PWN's magnetic field using the radio and X-ray
spectrum of the PWN.  In this paper, we evaluate the method described
by \citet{chevalier05}.  This procedure requires the radio and X-ray
spectral index of the PWN, the inferred break frequency $\nu_{\rm b}$,
and luminosity density of the PWN at the break frequency, $L_{\nu_{\rm
b}}$, between these bands, and assumes that $E_{\rm pwn,B} = (3/4)
E_{\rm pwn,p}$ (minimum energy estimate).  The values of $\nu_{\rm b}$
and $L_{\nu_{\rm b}}$ extrapolated from the radio and X-ray spectrum
of the PWN over-estimates $L_{\nu_{\rm b}}$ due to curvature in the
spectrum between these two bands (Figure \ref{photdensfig}).  This
fact, coupled with the assumption that $\sigma=3/7$ -- significantly
higher than the actual value of $\sigma$ (Figure \ref{sigmafig}),
causes this method to under-predict $E_{\rm pwn}$ by a factor of
$\sim5-10$ (Figure \ref{epwnminfig}) and over-predict $B_{\rm pwn}$ by
factors of a few at most times (Figure \ref{bpwnminfig}).

The above discussion assumes the PWN is not disrupted by any
hydrodynamical instabilities as it evolves inside the SNR.  However,
this is not necessarily the case.  The shell of swept-up material
surrounding the PWN is unstable to Rayleigh-Taylor instabilities when
$P_{\rm pwn}>P_{\rm snr}(R_{\rm pwn})$ because the low density pulsar
wind ($\rho_{\rm pwn}$) is accelerating this much higher density shell
($\rho_{\rm sw,pwn}$) \citep{chandrasekhar61}.  The growth rate of
these instabilities depends strongly on the magnetic field strength
inside the PWN parallel to the boundary between the pulsar wind and
swept-up supernova ejecta, $B_{\rm pwn,\parallel}$.  Numerical
simulations of the magnetic field inside PWN suggest that, at early
times, the PWN's magnetic field is largely toroidal
(e.g. \citealt{vdswaluw03}), in which case $B_{\rm pwn,\parallel}
\approx B_{\rm pwn}$ -- though polarized radio observations of older
PWNe (e.g. Vela X; \citealt{milne80}, \citealt{dodson03}) suggest a
strong radial component to their outer magnetic field.  Assuming
$B_{\rm pwn,\parallel}=B_{\rm pwn}$, the growth rate $\omega_{\rm
rt}(k)$ of a Rayleigh-Taylor (Kruskal-Schwarzschild;
\citealt{kruskal54}) instability with wavenumber $k\equiv2\pi/\lambda$
is \citep{chandrasekhar61,bucciantini04}:
\begin{equation}
\label{omegart}
\omega_{\rm rt}^2(k) = \frac{a_{\rm pwn}k(\rho_{\rm ms,pwn}-\rho_{\rm
    pwn})}{\rho_{\rm sw,pwn} + \rho_{\rm pwn}} - \frac{B_{\rm pwn}^2
    k^2}{2\pi(\rho_{\rm sw,pwn} + \rho_{\rm pwn})},
\end{equation}
where $a_{\rm pwn}$ is the acceleration of the shell of swept-up
material ($a_{\rm pwn}\equiv dv_{\rm pwn}/dt$).  Hydrodynamic
simulations of the expansion of the PWN inside a SNR suggest that this
shell of swept-up material that surrounds the PWN has thickness
$\approx \frac{1}{24}R_{\rm pwn}$ \citep{vdswaluw01b}, and we assume
this is true at all times when calculating $\rho_{\rm sw,pwn}$.  As a
result, the maximum wavenumber $k_{\rm crit}$ of a Rayleigh-Taylor
instability which can grow is \citep{bucciantini04}:
\begin{equation}
\label{kcriteqn}
k_{\rm crit} = \frac{2\pi a_{\rm pwn}}{B_{\rm pwn}^2}(\rho_{\rm
  ms,pwn} - \rho_{\rm pwn}),
\end{equation}
and the wavenumber of the Rayleigh-Taylor instability with the highest
growth rate $k_{\rm max}=k_{\rm crit}/2$
\citep{chandrasekhar61,stone07}.  Rayleigh-Taylor instabilities
between the pulsar wind and the swept-up ejecta result in ``bubbles''
of swept-up ejecta entering the PWN.  Three-dimensional numerical
simulations of the growth of Rayleigh-Taylor instabilities in a
PWN-like scenario (e.g., a strong magnetic field in the light fluid
parallel to the interface with the heavy fluid) suggest that the
penetration of bubbles is relatively unimpeded by the presence of a
magnetic field, though the magnetic field suppresses mixing between
these two fluids \citep{stone07}.  Therefore, to estimate the
penetration of swept-up ejecta bubbles into the PWN, we use the
results of relevant laboratory experiments.  These experiments derived
that the penetration depth of these bubbles, $h_{\rm rt}$, is
\citep{dimonte96,dimonte07}:
\begin{equation}
\label{hrteqn}
h_{\rm rt} = \alpha_b A \left[\int \sqrt{a_{\rm pwn}(t)} dt \right]^2,
\end{equation}
where $\alpha_b$ is an experimental derived constant ($\alpha_b \sim
0.061$; \citealt{dimonte96}), and $A$ is the Atwood number of this
system:
\begin{equation}
\label{atwood}
A \equiv \frac{\rho_{\rm ms,pwn} - \rho_{\rm pwn}}{\rho_{\rm ms,pwn} +
  \rho_{\rm pwn}}.
\end{equation}
Since $\rho_{\rm ms,pwn} \gg \rho_{\rm pwn}$, for a PWN $A\approx1$.

We find that the PWN is unstable to Rayleigh-Taylor instabilities
during the initial expansion ($t<t_{\rm col}$), and parts of the first
contraction, re-expansion ($t\sim15000-30000$ years), and second
contraction ($t\ga85000$ years).  During the initial expansion, we
expect that only Rayleigh-Taylor instabilities on the smallest angular
scales ($\la1^\circ$) are suppressed by the nebular magnetic field
(Figure \ref{rtscalesfig}), while during the first and second
contractions Rayleigh-Taylor instabilities at significantly larger
angular scale are suppressed (Figure \ref{rtscalesfig}) due to the
PWN's strong magnetic field (Figure \ref{bpwnfig}).  The growth rate
of these instabilities depends significantly on their angular scale,
and during the initial expansion instabilities with an angular size of
$\sim6^\circ$ grow the fastest (Figures \ref{rtscalesfig} \&
\ref{omegartfig}).  Initially during the contractions, the
Rayleigh-Taylor instabilities with the highest growth rates have an
angular scale $\sim5^\circ$ (Figure \ref{rtscalesfig} \&
\ref{omegartfig}).  However, as the PWN re-expands, the growth rate of
instabilities on these angular scales decreases and eventually are
suppressed -- for example, Rayleigh-Taylor instabilities with an
angular size of $\sim6^\circ$ can grow for only $\sim5000$ years
(Figure \ref{omegartfig}).  As a result, during this period
Rayleigh-Taylor instabilities with large angular scales
($\ga30^\circ$) likely experience the most growth (Figure
\ref{omegartfig}). Additionally, the growth-rate of Rayleigh-Taylor
instabilities at all angular scales is much lower after the PWN has
collided with the reverse shock than during the initial expansion due
to the stronger nebular magnetic field and significantly higher
density of the mass shell surrounding PWN (Figure \ref{omegartfig}).
Since the density of these perturbations grows as $\rho\propto
e^{\omega_{\rm rt}t}$ (\citealt{chandrasekhar61}), during the initial
expansion it is likely that a significant fraction of the mass
swept-up by the PWN will be in these filaments.

Rayleigh-Taylor instabilities will cause ``bubbles'' of swept-up
material to penetrate the PWN.  During the initial expansion of the
PWN, the depth of these bubbles is $h_{\rm rt}=(0.01-0.1)R_{\rm pwn}$
(Figure \ref{hrtfig}), so $\sim5-20\%$ of the volume of the PWN is
contained inside this penetration layer.  It is interesting to note
that this length scale is similar to the size of the optical filaments
surrounding the Crab Nebula \citep{hester96}.  After the PWN collides
with the reverse shock, Equation \ref{hrteqn} predicts that the depth
of this mixing layer decreases significantly, with $h_{\rm
rt}\approx0$ when the PWN begins to contract (Figure \ref{hrtfig}).
This mixing layer will grow again when the pressure inside the PWN is
higher than that of the surrounding SNR.  When the PWN re-expands into
the SNR, this mixing layer is almost as large as the PWN itself
(Figure \ref{hrtfig}).  This suggests tat the PWN might be disrupted
at this time, as observed in previous hydrodynamical simulations
\citep{blondin01, vdswaluw04}.  If so, this will inject a total energy
of $\sim6\times10^{48}$ ergs in the form of relativistic particles
into the surrounding SNR.  It is important to reiterate that the
calculation of $h_{\rm rt}$ used here (Equation \ref{hrteqn}) ignores
any damping the PWN's magnetic field might have on the growth of this
maxing layer -- as do the hydrodynamical simulations cited above.
Recent simulations of the growth of Rayleigh-Taylor instabilities
inside a PWN suggest the growth of this mixing layer is highly
suppressed for high $\sigma$ \citep{bucciantini04}.  Since this is the
case during the contraction of the PWN (Figure \ref{sigmafig}), it is
likely this approach under-predicts the lifetime of a PWN inside a
SNR.

\section{Discussion and Conclusions}
\label{discussion}

In this paper, we present a general model for the evolution of a PWN
inside a SNR (\S\ref{theory}) and the specific evolution predicted
this model for a particular set of neutron star, pulsar wind,
supernova explosion, and ISM properties (\S\ref{performance}).  As
mentioned in \S\ref{intro}, the ultimate goal of this model is to
reproduce the observed dynamical and radiative properties of a PWN in
order to study the central neutron star, progenitor supernova
explosion, and pulsar wind.  This requires that our model accurately
reproduces the observed properties of a well-studied and constrained
PWN.  The best test case is the Crab Nebula, the brightest radio and
X-ray PWN in the Milky Way and whose age ($\approx950$ years old) and
neutron star properties ($\dot{E}_0$, $p$, $\tau_{\rm sd}$) are well
known.  As mentioned earlier, the input parameters of the simulation
discussed in \S\ref{performance} were based on previous analyses of
this source.  As shown in Table \ref{crabneb}, this model is able to
reproduce the size of the Crab Nebula, the radius of the termination
shock, the expansion velocity of the PWN, and the spectral index of
both the radio and 0.5--10 keV emission from the PWN.  However, for
the set of input parameters given in Table \ref{trialmodel}, this
model predicts a radio, X-ray, and TeV $\gamma$-ray luminosity
$\sim10\times$ different than that observed.  Since the relationship
between the properties of a PWN at a given time and the input
parameters of this model is non-trivial, determining if this model can
reproduce all of the observed properties of the Crab Nebula requires a
much more thorough examination of the possible parameter space, which
we leave for future work.  This situation is true for any PWN, and not
just the Crab Nebula.

The evolution of the particular PWN discussed in \S\ref{performance}
does have some interesting implications for existing questions in the
field.  The spectrum of this PWN does show the large spectral breaks
inferred from radio and X-ray observations of several young PWNe
though, though not their low break frequency \citep{woltjer97}.  By
exploring the parameter space of possible neutron star, pulsar wind,
supernova, and ISM properties, we will be able to identify what
regimes are necessary to produce a low break frequency and large
spectral break between the radio and X-ray wavebands at early times.
For the set of parameters presented in \S\ref{performance}, this model
predicts a very low break frequency ($\nu\sim1$ GHz) at late times --
as observed in some older ($t\ga10^4$ years) PWNe (e.g. DA 495;
\citealt{kothes08}).  This model also predicts that this particular
PWN will have a low radio luminosity but a high TeV $\gamma$-ray
luminosity at late times, similar to several PWN recently discovered
by HESS (e.g. \citealt{aharonian06}).  Additionally, this model that
-- for this set of parameters -- the magnetization of the PWN
$\sigma$ is considerably higher after the PWN collides with the
reverse shock than before.  This could explain why the value of
$\sigma$ estimated for the Vela PWN ($0.05<\sigma<0.5$ ;
\citealt{sefako03}), a system where this collision is believed to have
already occurred (e.g. \citealt{lamassa08}), is considerably higher
than that of the Crab Nebula ($\sigma\sim0.003$; \citealt{kennel84}),
where this is not the case.  Finally, at these late times the spectrum
of the PWN discussed in \S\ref{performance} has several features at
$\gamma$-ray and low frequency ($\nu<1$ GHz) radio wavelengths which
are observed with new facilities such as the {\it Fermi Gamma-ray
Space Telescope}, LOFAR, LWA, MWA, and EVLA.  Parameter exploration is
required to determine how general these conclusions are, but suggest
this model will be useful in resolving several outstanding problems
concerning the evolution of a PWN inside a SNR.

Additionally, this model is extremely useful in determining if PWNe
inside SNRs can deposit a sufficient number of high energy electrons
and positrons into the surrounding ISM to explain the rising positron
fraction of cosmic-ray leptons detected by {\it PAMELA} between
$\sim1.5-100$ GeV \citep{adriani09} and the excess of cosmic ray
electrons and positrons detected by ATIC \citep{chang08} and HESS
\citep{aharonian08} between $\sim300-800$ GeV.  These results require
a nearby source of high-energy electrons and positrons, and one
possibility is PWNe inside SNRs (e.g. \citealt{malyshev09}).  If this
is correct, the average PWN must deposit $\sim10^{49}$ ergs of
energetic electrons and positrons into the surrounding ISM, and these
particles must have an energy spectrum flatter than $E^{-2}$ which
extends up to an energy of $\sim1$ TeV \citep{malyshev09}.  For the
set of parameters modeled in \S\ref{performance}, these conditions are
met for only a short period of time during the evolution of this PWN.
Using this model, it is possible to determine what sets of neutron
star, pulsar wind, supernova, and ISM parameters are required for the
PWN to satisfy these criteria for a longer period of time, and
evaluate different models for particle escape from the PWN and their
effect on the PWN's evolution -- particularly if these particles
escape gradually or suddenly from the PWN.

One major limitation of this model is that is in inherently
one-dimensional since it assumes a constant pressure and magnetic
field strength inside the PWN.  Therefore, it is insensitive to any
pressure and magnetic field variations inside the PWN.  Recent {\it
Chandra} observations of PWNe have revealed the existence of internal
structure (e.g., the torus and jets in the Crab Nebula;
\citealt{weisskopf00}) but of spectral changes as well
\citep{weisskopf00, mori04} -- indicative of a non-uniform pressure
and/or magnetic field inside the PWN.  We are also insensitive to
possibly significant effects asymmetries in the PWN resulting from
either the space velocity of the neutron star or inhomogeneities in
the progenitor supernova and/or surrounding ISM which can have on its
evolution, particularly after it collides with the SNR reverse shock
(e.g. \citealt{vdswaluw04}).  Additionally, while we estimate the growth
rate of Rayleigh-Taylor instabilities in the shell of swept-up
material surrounding the PWN, we can not determine the role they might
play in enabling particles to escape from the PWN -- critical in
determining if such PWNe are responsible for the {\it PAMELA}, ATIC,
and HESS results discussed above.

To summarize, we present a one-dimensional model for the evolution of
a PWN inside a SNR.  This model self-consistently evolves the
magnetic, dynamical, and radiative properties of the PWN throughout
its evolution, and therefore represents a significant improvement over
other currently existing models.  The model described here provides a
framework for investigating the effect of more complicated
descriptions of pulsar winds, e.g. the presence of ions in the pulsar
wind, on the dynamical and radiative evolution of the PWN, as well as
determining if there exists any unique observational signatures of
such processes.  Additionally, it is well-suited for using the
observed properties of a PWN to constrain the properties of the
central neutron star, its pulsar wind, progenitor supernova, and
surrounding ISM.  Its applicability to a large number of known,
well-studied PWNe make it is an extremely powerful for studying these
intriguing systems.


\acknowledgements We would like to thank Roger Chevalier, Andrei
Gruzinov, Andrew MacFadyen, and Matias Zaldariagga for useful
discussions.  JDG is supported by an NSF Astronomy and Astrophysics
Postdoctoral Fellowship under award AST-0702957.

\clearpage
\begin{table*}
\caption{Model Input Parameters \label{modelinput}}
\begin{center}
\begin{tabular}{ccc}
\hline
\hline
{\sc Parameter} & {\sc Units} & {\sc Description} \\
\hline
\multicolumn{3}{c}{\it Supernova Explosion} \\
$E_{\rm sn,51}$ & $10^{51}$ ergs & Initial Kinetic Energy of Supernova
Ejecta \\
$M_{\rm ej}$ & $M_{\odot}$ & Mass of Supernova Ejecta \\
\hline
\multicolumn{3}{c}{\it Interstellar Medium} \\
$n_{\rm ism}$ & cm$^{-3}$ & Number density of surrounding ISM \\
\hline
\multicolumn{3}{c}{\it Neutron Star} \\
$p$ & $\cdots$ & Neutron Star Braking Index \\
$\tau_{\rm sd}$ & years & Characteristic Spin-down Timescale of the
Neutron Star \\
$\dot{E}_{0,40}$ & $10^{40}$ ergs s$^{-1}$ & Initial Spin-down
Luminosity of the Neutron Star \\
$v_{\rm psr}$ & km s$^{-1}$ & Space velocity of the Neutron Star \\
\hline
\multicolumn{3}{c}{\it Pulsar Wind} \\
$\eta_e$ & $\cdots$ & Fraction of neutron's star spin-down luminosity
injected as electrons \\
$\eta_i$ & $\cdots$ & Fraction of neutron's star spin-down luminosity
injected as ions \\
$\eta_{\rm B}$ & $\cdots$ & Fraction of neutron's star spin-down luminosity
injected as magnetic energy \\
$E_{\rm e,min}$ & varies & Minimum Energy of Electrons injected into the
PWN \\
$E_{\rm e,max}$ & varies & Maximum Energy of Electrons injected into the
PWN \\
$E_{\rm i,min}$ & varies & Minimum Energy of Ions injected into the PWN \\
$E_{\rm i,max}$ & varies & Maximum Energy of Ions injected into the PWN \\
$\gamma_e$ & $\cdots$ & Electron injection index \\
$\gamma_i$ & $\cdots$ & Ion injection index \\
\hline
\hline
\end{tabular}
\end{center}
\end{table*}

\begin{table*}
\begin{center}
\caption{Trial Model Input Parameters \label{trialmodel}}
\begin{tabular}{cc}
\hline
\hline
{\sc Parameter} & {\sc Value} \\
\hline
$E_{\rm sn,51}$ & 1 \\
$M_{\rm ej}$ & 8 \\
$n_{\rm ism}$ & 0.1 \\
$p$ & 3.00 \\
$\tau_{\rm sd}$ & 500 \\ 
$\dot{E}_{0,40}$ & 1 \\
$v_{\rm psr}$ & 120 \\ 
$\eta_e$ & 0.999 \\
$\eta_i$ & 0 \\
$\eta_{\rm B}$ & 0.001 \\
$E_{\rm e,min}$ & 511 keV \\
$E_{\rm e,max}$ & 500 TeV \\
$E_{\rm i,min}$ & $\cdots$ \\
$E_{\rm i,max}$ & $\cdots$ \\
$\gamma_e$ & 1.6 \\
$\gamma_i$ & $\cdots$ \\
\hline
\hline
\end{tabular}
\end{center}
\end{table*}

\begin{table*}
\begin{center}
\caption{Selected Observed Properties of the Crab Nebula \label{crabneb}}
\begin{tabular}{ccc|c}
\hline
\hline
{\sc Observed Property} & {\sc Value} & {\sc Reference} & {\sc Model
  Prediction} \\
\hline
$R_{\rm pwn}$ & 1.5-2 pc & \citet{green06} & 1.7 pc \\
$r_{\rm ts}$ & 0.07-0.14 pc & \citet{weisskopf00} & 0.24 pc \\
$v_{\rm pwn}$ & $\sim1270$ km s$^{-1}$ & \citet{temim06} & 2000 km
s$^{-1}$ \\
Radio Luminosity & $1.8\times10^{35}$ ergs s$^{-1}$ & \citet{frail97}
& $2.2\times10^{34}$ ergs s$^{-1}$ \\
$\alpha_{\rm radio}$ & $-0.3$ & \citet{green06} & $-0.3$ \\
$L_{\rm X,0.5-10 keV}$ & $1.3\times10^{37}$ ergs s$^{-1}$ &
\citet{mori04} & $\sim3\times10^{38}$ ergs s$^{-1}$ \\
$\Gamma_{\rm 0.5-10 keV}$ & 1.9-3.0 & \citet{mori04} & 1.8 \\
$L_{\rm 50 GeV-50 TeV}$ & $\sim10^{34}-10^{35}$ ergs s$^{-1}$ &
\citet{aharonian04} & $1.6\times10^{36}$ ergs s$^{-1}$ \\

\hline
\hline
\end{tabular}
\end{center}
\end{table*}

\onecolumn
\clearpage
\begin{figure}
\begin{center}
\end{center}
\caption{The ratio of the inverse Compton cross-section of an electron
  scattering off the CMB relative to the Thomson cross-section.}
\label{powicratfig}
\end{figure}

\begin{figure}
\begin{center}
\end{center}
\caption{Radius of the SNR ($R_{\rm snr}$, blue line), reverse shock
  ($R_{\rm rs}$, green line), PWN ($R_{\rm pwn}$, red line), the
  position of the neutron star ($r_{\rm psr}$, orange line), and the
  radius of the termination shock inside the PWN ($r_{\rm ts}$, purple
  line) for a neutron star, supernova, and PWN with the properties
  listed in Table \ref{trialmodel}.  The termination shock only exists
  when the pulsar is inside the PWN, and is always centered on the
  pulsar.  The vertical dotted lines indicate the division between the
  four evolutionary phases of the PWN discussed in \S\ref{performance}.}
\label{radpwnfig}
\end{figure}

\begin{figure}
\begin{center}
\end{center}
\caption{The pressure inside the PWN ($P_{\rm pwn}$, red) and the
  pressure inside the SNR just outside the PWN ($P_{\rm snr}(R_{\rm
  pwn})$, blue).  At times $t<t_{\rm col}$, we assume that $P_{\rm
  snr}(R_{\rm pwn})=0$.  The vertical dotted lines indicate the
  division between the four evolutionary phases of the PWN discussed
  in \S\ref{performance}.}
\label{pressfig}
\end{figure}

\begin{figure}
\begin{center}
\end{center}
\caption{The velocity of the shell of swept-up material surrounding
  the PWN ($v_{\rm pwn}$, red) and the velocity of the supernova
  ejecta just beyond this mass shell [$v_{\rm ej}(R_{\rm pwn})$,
  blue].  The vertical dotted lines indicate the division between the
  four evolutionary phases of the PWN discussed in \S\ref{performance}.}
\label{velfig}
\end{figure}

\begin{figure}
\begin{center}
\end{center}
\caption{The mass of material swept-up by the PWN ($M_{\rm sw,pwn}$)
  as it expands inside the SNR.  The vertical dotted lines indicate
  the division between the four evolutionary phases of the PWN
  discussed in \S\ref{performance}.}
\label{mswfig}
\end{figure}

\begin{figure}
\begin{center}
\end{center}
\caption{The spin-down luminosity of the pulsar ($L_{\rm
  psr}\equiv\dot{E}$, blue) and the adiabatic ($L_{\rm ad}$, green),
  synchrotron ($L_{\rm synch}$, red), and inverse Compton ($L_{\rm
  IC}$, orange) luminosity of the PWN.  Solid lines indicate energy
  lost by the PWN, while dashed lines indicate energy gained by the
  PWN.  The vertical dotted lines indicate the division between the
  four evolutionary phases of the PWN discussed in
  \S\ref{performance}.}
\label{edotfig}
\end{figure}

\begin{figure}
\begin{center}
\end{center}
\caption{The total energy ($E_{\rm pwn}$, red), particle energy
  ($E_{\rm pwn,p}$, green) and magnetic energy ($E_{\rm pwn,B}$, blue)
  of the PWN. The vertical dotted lines indicate the division between the
  four evolutionary phases of the PWN discussed in \S\ref{performance}.} 
\label{epwnfig}
\end{figure}

\clearpage
\begin{figure}
\begin{center}
\end{center}
\caption{Magnetic field strength of the PWN, $B_{\rm pwn}$. The
  vertical dotted lines indicate the division between the four
  evolutionary phases of the PWN discussed in \S\ref{performance}.}
\label{bpwnfig}
\end{figure}

\begin{figure}
\begin{center}
\end{center}
\caption{The magnetization parameter of the PWN $\sigma$ for $t<t_{\rm
  col}$ ({\it top}) and $t>t_{\rm col}$ ({\it bottom}).  In the bottom
  plot, the vertical lines represent the division between the second,
  third, and fourth evolution phase of the PWN discussed in
  \S\ref{performance}.}
\label{sigmafig}
\end{figure}

\begin{figure}
\begin{center}
\end{center}
\caption{The energy spectrum of electrons and positrons inside the PWN
  during the first ({\it upper left}), second ({\it upper right}),
  third ({\it bottom left}), and fourth ({\it bottom right})
  evolutionary phase discussed in \S\ref{performance}.  In all four
  plots, the different color lines correspond to the energy spectrum
  at different ages, the vertical dotted lines indicate $E_{\rm
  e,min}$ ({\it right}) and $E_{\rm e,max}$ ({\it left}), and the
  dashed line at the bottom of the plot indicates the shape and extent
  of the injected spectrum.  In the {\it upper-right} plot, the black
  line corresponds to the energy spectrum at the time the pulsar
  leaves its PWN for the first time.  In the {\it bottom left} plot,
  the dotted orange line corresponds to the energy spectrum at the
  time when the pulsar re-enters the PWN.  In the {\it bottom right}
  plot, the dotted yellow line corresponds to the energy spectrum just
  before the pulsar re-enters the PWN.}
\label{elecspec}
\end{figure}

\begin{figure}
\begin{center}
\end{center}
\caption{The spectrum of photons radiated by the PWN during the first
  ({\it upper left}), second ({\it upper right}), third ({\it bottom
  left}), and fourth ({\it bottom right}) evolutionary phase discussed
  in \S\ref{performance}.  In all four plots, the different color
  lines correspond to the photon spectrum at different ages, and the
  cross-hatched regions indicate, from {\it left} to {\it right}, the
  radio, mid-infrared, optical, soft X-ray, hard X-ray, $\gamma$-ray,
  and TeV $\gamma$-ray regimes of the electromagnetic spectrum.  In
  the {\it upper-right} plot, the black line corresponds to the photon
  spectrum at the time the pulsar leaves its PWN for the first time.
  In the {\it bottom left} plot, the dotted orange line corresponds to
  the photon spectrum at the time when the pulsar re-enters the PWN.
  In the {\it bottom right} plot, the dotted yellow line corresponds
  to the photon spectrum just before the pulsar re-enters the PWN.}
\label{photspec}
\end{figure}

\begin{figure}
\begin{center}
\end{center}
\caption{The photon luminosity of the PWN in the radio
  ($\nu=10^7-10^{11}$ Hz, $L_{\rm radio}$, red), mid-infrared
  ($\lambda=3.6-160\mu$m, orange), near-infrared/optical
  ($\lambda=2.35\mu$m--354.3 nm, yellow), soft X-ray ($h\nu=0.5-10$
  keV, dark green), hard X-ray ($h\nu=$15 keV--10 MeV, blue),
  $\gamma$-ray ($h\nu=10$ MeV--100 GeV, dark blue), and TeV
  $\gamma$-ray ($h\nu=50$ GeV -- 50 TeV, purple) for $t\leq t_{\rm
  col}$ ({\it top}) and $t>t_{\rm col}$ ({\it bottom}).  The
  definition of the wavebands were chosen to reflect the frequency /
  wavelength / energy coverage of current observing facilities, and
  are given in units usually associated with that portion of the
  electromagnetic spectrum.  In the bottom plot, the vertical dotted
  lines demarcate the second, third, and fourth evolutionary phase
  discussed in \S\ref{performance}.}
\label{photlumin}
\end{figure}

\begin{figure}
\begin{center}
\end{center}
\caption{The spectral index $\alpha$ (photon index $\Gamma$) of the
  radio (red), 0.5--10 keV (green), and 2--10 keV (blue) emission from
  the PWN.  The dotted vertical lines indicate the evolutionary phases
  discussed in \S\ref{performance}.}
\label{alphafig}
\end{figure}

\begin{figure}
\begin{center}
\end{center}
\caption{The luminosity density of the PWN between the radio and soft
  X-ray bands during the first ({\it upper left}), second ({\it upper
  right}), third ({\it bottom left}), and fourth ({\it bottom right})
  evolutionary phase discussed in \S\ref{performance}.  The filled and
  open stars indicate $\nu_{\rm b}$ and $L_{\nu_b}$ derived by joining
  power law fits to the radio and 0.5--10 keV (closed stars) or 2--10
  keV (open stars) spectrum of the PWN.}
\label{photdensfig}
\end{figure}

\begin{figure}
\begin{center}
\end{center}
\caption{The change in spectral index $\Delta\alpha$ between the radio
  and 0.5--10 keV band (green) and the radio and 2--10 keV band
  (blue).  The dotted vertical lines indicate the evolutionary phases
  discussed in \S\ref{performance}.}
\label{delalphafig}
\end{figure}

\begin{figure}
\begin{center}
\end{center}
\caption{The total energy inside the PWN ($E_{\rm pwn}$, red) and the
  minimum energy inside the PWN calculated using the radio and 0.5--10
  keV spectrum of the PWN ($E_{\rm pwn,min (0.5-10 keV}$, blue) and
  the radio and 2--10 keV spectrum of the PWN ($E_{\rm pwn,min (0.5-10
  keV}$, green).  The dotted vertical lines indicate the evolutionary
  phases discussed in \S\ref{performance}.}
\label{epwnminfig}
\end{figure}

\begin{figure}
\begin{center}
\end{center}
\caption{Magnetic field strength of the PWN, $B_{\rm pwn}$ (red), as
  well as the magnetic field strength inferred using the radio and
  0.5--10 keV spectrum of the PWN ($B_{\rm pwn,min (0.5-10 keV)}$,
  blue) and the radio and 2--10 keV spectrum of the PWN ($B_{\rm
  pwn,min (2-10 keV)}$, green).  The dotted vertical lines indicate
  the evolutionary phases discussed in \S\ref{performance}.}
\label{bpwnminfig}
\end{figure}

\begin{figure}
\begin{center}
\end{center}
\caption{The minimum angular scale unstable to Rayleigh-Taylor
  instabilities ($\theta_{\rm rt,crit}\equiv\lambda_{\rm
  rt,crit}/R_{\rm pwn}$; red) and the angular scale maximally unstable
  to Rayleigh-Taylor instabilities ($\theta_{\rm rt,max} \equiv
  \lambda_{\rm rt,max}/R_{\rm pwn}$; blue) for $t\leq t_{\rm col}$
  ({\it top}) and $t>t_{\rm col}$ ({\it bottom}). In the bottom plot,
  the vertical dotted lines demarcate the second, third, and fourth
  evolutionary phase discussed in \S\ref{performance}.}
\label{rtscalesfig}
\end{figure}

\begin{figure}
\begin{center}
\end{center}
\caption{The growth rate ($\omega_{\rm rt}$, Equation \ref{omegart})
  of Rayleigh-Taylor instabilities with different angular scales for
  $t\leq t_{\rm col}$ ({\it top}) and $t>t_{\rm col}$ ({\it
  bottom}). In the bottom plot, the vertical dotted lines demarcate
  the second, third, and fourth evolutionary phase discussed in
  \S\ref{performance}.}
\label{omegartfig}
\end{figure}

\begin{figure}
\begin{center}
\end{center}
\caption{The penetration depth of Rayleigh-Taylor bubbles into the PWN
  ($h_{\rm rt}$, orange) and radius of the PWN ($R_{\rm pwn}$, red).
  The dotted vertical lines indicate the evolutionary phases discussed
  in \S\ref{performance}.}
\label{hrtfig}
\end{figure}

\clearpage
\bibliographystyle{apj} \bibliography{ms} 

\begin{thebibliography}{62}
\expandafter\ifx\csname natexlab\endcsname\relax\def\natexlab#1{#1}\fi

\bibitem[{{Adriani} {et~al.}(2009){Adriani}, {Barbarino}, {Bazilevskaya},
  {Bellotti}, {Boezio}, {Bogomolov}, {Bonechi}, {Bongi}, {Bonvicini}, {Bottai},
  {Bruno}, {Cafagna}, {Campana}, {Carlson}, {Casolino}, {Castellini}, {de
  Pascale}, {de Rosa}, {de Simone}, {di Felice}, {Galper}, {Grishantseva},
  {Hofverberg}, {Koldashov}, {Krutkov}, {Kvashnin}, {Leonov}, {Malvezzi},
  {Marcelli}, {Menn}, {Mikhailov}, {Mocchiutti}, {Orsi}, {Osteria}, {Papini},
  {Pearce}, {Picozza}, {Ricci}, {Ricciarini}, {Simon}, {Sparvoli},
  {Spillantini}, {Stozhkov}, {Vacchi}, {Vannuccini}, {Vasilyev}, {Voronov},
  {Yurkin}, {Zampa}, {Zampa}, \& {Zverev}}]{adriani09}
{Adriani} et al. 2009, \nat, 458, 607

\bibitem[{{Aharonian} {et~al.}(2004){Aharonian}, {Akhperjanian}, {Beilicke},
  {Bernl{\"o}hr}, {B{\"o}rst}, {Bojahr}, {Bolz}, {Coarasa}, {Contreras},
  {Cortina}, {Denninghoff}, {Fonseca}, {Girma}, {G{\"o}tting}, {Heinzelmann},
  {Hermann}, {Heusler}, {Hofmann}, {Horns}, {Jung}, {Kankanyan}, {Kestel},
  {Kohnle}, {Konopelko}, {Kranich}, {Lampeitl}, {Lopez}, {Lorenz}, {Lucarelli},
  {Mang}, {Mazin}, {Meyer}, {Mirzoyan}, {Moralejo}, {O{\~n}a-Wilhelmi},
  {Panter}, {Plyasheshnikov}, {P{\"u}hlhofer}, {de los Reyes}, {Rhode},
  {Ripken}, {Rowell}, {Sahakian}, {Samorski}, {Schilling}, {Siems},
  {Sobzynska}, {Stamm}, {Tluczykont}, {Vitale}, {V{\"o}lk}, {Wiedner}, \&
  {Wittek}}]{aharonian04}
{Aharonian} et al. 2004, \apj, 614, 897

\bibitem[{{Aharonian} {et~al.}(2008){Aharonian}, {Akhperjanian}, {Barres de
  Almeida}, {Bazer-Bachi}, {Becherini}, {Behera}, {Benbow}, {Bernl{\"o}hr},
  {Boisson}, {Bochow}, {Borrel}, {Braun}, {Brion}, {Brucker}, {Brun},
  {B{\"u}hler}, {Bulik}, {B{\"u}sching}, {Boutelier}, {Carrigan}, {Chadwick},
  {Charbonnier}, {Chaves}, {Cheesebrough}, {Chounet}, {Clapson}, {Coignet},
  {Costamante}, {Dalton}, {Degrange}, {Deil}, {Dickinson}, {Djannati-Ata{\"i}},
  {Domainko}, {Drury}, {Dubois}, {Dubus}, {Dyks}, {Dyrda}, {Egberts},
  {Emmanoulopoulos}, {Espigat}, {Farnier}, {Feinstein}, {Fiasson},
  {F{\"o}rster}, {Fontaine}, {F{\"u}{\ss}ling}, {Gabici}, {Gallant},
  {G{\'e}rard}, {Giebels}, {Glicenstein}, {Gl{\"u}ck}, {Goret},
  {Hadjichristidis}, {Hauser}, {Hauser}, {Heinz}, {Heinzelmann}, {Henri},
  {Hermann}, {Hinton}, {Hoffmann}, {Hofmann}, {Holleran}, {Hoppe}, {Horns},
  {Jacholkowska}, {de Jager}, {Jung}, {Katarzy{\'n}ski}, {Kaufmann},
  {Kendziorra}, {Kerschhaggl}, {Khangulyan}, {Kh{\'e}lifi}, {Keogh}, {Komin},
  {Kosack}, {Lamanna}, {Lenain}, {Lohse}, {Marandon}, {Martin},
  {Martineau-Huynh}, {Marcowith}, {Maurin}, {McComb}, {Medina}, {Moderski},
  {Moulin}, {Naumann-Godo}, {de Naurois}, {Nedbal}, {Nekrassov}, {Niemiec},
  {Nolan}, {Ohm}, {Olive}, {de O{\~n}a Wilhelmi}, {Orford}, {Osborne},
  {Ostrowski}, {Panter}, {Pedaletti}, {Pelletier}, {Petrucci}, {Pita},
  {P{\"u}hlhofer}, {Punch}, {Quirrenbach}, {Raubenheimer}, {Raue}, {Rayner},
  {Renaud}, {Rieger}, {Ripken}, {Rob}, {Rosier-Lees}, {Rowell}, {Rudak},
  {Rulten}, {Ruppel}, {Sahakian}, {Santangelo}, {Schlickeiser}, {Sch{\"o}ck},
  {Schr{\"o}der}, {Schwanke}, {Schwarzburg}, {Schwemmer}, {Shalchi}, {Skilton},
  {Sol}, {Spangler}, {Stawarz}, {Steenkamp}, {Stegmann}, {Superina}, {Tam},
  {Tavernet}, {Terrier}, {Tibolla}, {van Eldik}, {Vasileiadis}, {Venter},
  {Vialle}, {Vincent}, {Vivier}, {V{\"o}lk}, {Volpe}, {Wagner}, {Ward},
  {Zdziarski}, \& {Zech}}]{aharonian08}
{Aharonian} et al. 2008, Physical Review Letters, 101, 261104

\bibitem[{{Aharonian} {et~al.}(2006){Aharonian}, {Akhperjanian}, {Bazer-Bachi},
  {Beilicke}, {Benbow}, {Berge}, {Bernl{\"o}hr}, {Boisson}, {Bolz}, {Borrel},
  {Braun}, {Breitling}, {Brown}, {Chadwick}, {Chounet}, {Cornils},
  {Costamante}, {Degrange}, {Dickinson}, {Djannati-Ata{\"i}}, {Drury}, {Dubus},
  {Emmanoulopoulos}, {Espigat}, {Feinstein}, {Fontaine}, {Fuchs}, {Funk},
  {Gallant}, {Giebels}, {Gillessen}, {Glicenstein}, {Goret}, {Hadjichristidis},
  {Hauser}, {Heinzelmann}, {Henri}, {Hermann}, {Hinton}, {Hofmann}, {Holleran},
  {Horns}, {Jacholkowska}, {de Jager}, {Kh{\'e}lifi}, {Komin}, {Konopelko},
  {Latham}, {Le Gallou}, {Lemi{\`e}re}, {Lemoine-Goumard}, {Leroy}, {Lohse},
  {Martin}, {Martineau-Huynh}, {Marcowith}, {Masterson}, {McComb}, {de
  Naurois}, {Nolan}, {Noutsos}, {Orford}, {Osborne}, {Ouchrif}, {Panter},
  {Pelletier}, {Pita}, {P{\"u}hlhofer}, {Punch}, {Raubenheimer}, {Raue},
  {Raux}, {Rayner}, {Reimer}, {Reimer}, {Ripken}, {Rob}, {Rolland}, {Rowell},
  {Sahakian}, {Saug{\'e}}, {Schlenker}, {Schlickeiser}, {Schuster}, {Schwanke},
  {Siewert}, {Sol}, {Spangler}, {Steenkamp}, {Stegmann}, {Tavernet}, {Terrier},
  {Th{\'e}oret}, {Tluczykont}, {Vasileiadis}, {Venter}, {Vincent}, {V{\"o}lk},
  \& {Wagner}}]{aharonian06}
{Aharonian} et al. 2006, \apj, 636, 777

\bibitem[{{Arons}(2007)}]{arons07}
{Arons}, J. 2007, ArXiv e-prints (astro-ph/0708.1050)

\bibitem[{Baade \& Zwicky(1934)}]{baade34}
Baade, W. \& Zwicky, F. 1934, Phys. Rev., 46, 76

\bibitem[{{Bandiera}(1984)}]{bandiera84}
{Bandiera}, R. 1984, \aap, 139, 368

\bibitem[{{Blondin} {et~al.}(2001){Blondin}, {Chevalier}, \&
  {Frierson}}]{blondin01}
{Blondin}, J.~M., {Chevalier}, R.~A., \& {Frierson}, D.~M. 2001, \apj, 563, 806

\bibitem[{{Blondin} {et~al.}(1998){Blondin}, {Wright}, {Borkowski}, \&
  {Reynolds}}]{blondin98}
{Blondin}, J.~M., {Wright}, E.~B., {Borkowski}, K.~J., \& {Reynolds}, S.~P.
  1998, \apj, 500, 342

\bibitem[{{Bucciantini} {et~al.}(2004){Bucciantini}, {Amato}, {Bandiera},
  {Blondin}, \& {Del Zanna}}]{bucciantini04}
{Bucciantini}, N., {Amato}, E., {Bandiera}, R., {Blondin}, J.~M., \& {Del
  Zanna}, L. 2004, \aap, 423, 253

\bibitem[{{Bucciantini} {et~al.}(2003){Bucciantini}, {Blondin}, {Del Zanna}, \&
  {Amato}}]{bucciantini03}
{Bucciantini}, N., {Blondin}, J.~M., {Del Zanna}, L., \& {Amato}, E. 2003,
  \aap, 405, 617

\bibitem[{{Chandrasekhar}(1961)}]{chandrasekhar61}
{Chandrasekhar}, S. 1961, {Hydrodynamic and hydromagnetic stability}
  (International Series of Monographs on Physics, Oxford: Clarendon, 1961)

\bibitem[{{Chang} {et~al.}(2008){Chang}, {Adams}, {Ahn}, {Bashindzhagyan},
  {Christl}, {Ganel}, {Guzik}, {Isbert}, {Kim}, {Kuznetsov}, {Panasyuk},
  {Panov}, {Schmidt}, {Seo}, {Sokolskaya}, {Watts}, {Wefel}, {Wu}, \&
  {Zatsepin}}]{chang08}
{Chang} et al. 2008, \nat, 456, 362

\bibitem[{{Chevalier}(1982)}]{chevalier82}
{Chevalier}, R.~A. 1982, \apj, 258, 790

\bibitem[{{Chevalier}(2005)}]{chevalier05}
---. 2005, \apj, 619, 839

\bibitem[{{Chevalier} \& {Fransson}(1992)}]{chevalier92}
{Chevalier}, R.~A. \& {Fransson}, C. 1992, \apj, 395, 540

\bibitem[{{Del Zanna} {et~al.}(2004){Del Zanna}, {Amato}, \&
  {Bucciantini}}]{delzanna04}
{Del Zanna}, L., {Amato}, E., \& {Bucciantini}, N. 2004, \aap, 421, 1063

\bibitem[{{Dimonte} {et~al.}(2007){Dimonte}, {Ramaprabhu}, \&
  {Andrews}}]{dimonte07}
{Dimonte}, G., {Ramaprabhu}, P., \& {Andrews}, M. 2007, \pre, 76, 046313

\bibitem[{{Dimonte} \& {Schneider}(1996)}]{dimonte96}
{Dimonte}, G. \& {Schneider}, M. 1996, \pre, 54, 3740

\bibitem[{{Dodson} {et~al.}(2003){Dodson}, {Lewis}, {McConnell}, \&
  {Deshpande}}]{dodson03}
{Dodson}, R., {Lewis}, D., {McConnell}, D., \& {Deshpande}, A.~A. 2003, \mnras,
  343, 116

\bibitem[{{Frail} \& {Scharringhausen}(1997)}]{frail97}
{Frail}, D.~A. \& {Scharringhausen}, B.~R. 1997, \apj, 480, 364

\bibitem[{{Gaensler} \& {Slane}(2006)}]{gaensler06}
{Gaensler}, B.~M. \& {Slane}, P.~O. 2006, \araa, 44, 17

\bibitem[{{Gelfand} {et~al.}(2007){Gelfand}, {Gaensler}, {Slane}, {Patnaude},
  {Hughes}, \& {Camilo}}]{gelfand07}
{Gelfand}, J.~D., {Gaensler}, B.~M., {Slane}, P.~O., {Patnaude}, D.~J.,
  {Hughes}, J.~P., \& {Camilo}, F. 2007, \apj, 663, 468

\bibitem[{{Goldreich} \& {Julian}(1969)}]{goldreich69}
{Goldreich}, P. \& {Julian}, W.~H. 1969, \apj, 157, 869

\bibitem[{{Green}(2006)}]{green06}
{Green}, D.~A. 2006, VizieR Online Data Catalog, 7227, 0

\bibitem[{{Hester} {et~al.}(1996){Hester}, {Stone}, {Scowen}, {Jun},
  {Gallagher}, {Norman}, {Ballester}, {Burrows}, {Casertano}, {Clarke},
  {Crisp}, {Griffiths}, {Hoessel}, {Holtzman}, {Krist}, {Mould}, {Sankrit},
  {Stapelfeldt}, {Trauger}, {Watson}, \& {Westphal}}]{hester96}
{Hester}, J.~J et al. 1996, \apj, 456, 225

\bibitem[{{Jun}(1998)}]{jun98}
{Jun}, B.-I. 1998, \apj, 499, 282

\bibitem[{{Kaplan} {et~al.}(2008){Kaplan}, {Chatterjee}, {Gaensler}, \&
  {Anderson}}]{kaplan08}
{Kaplan}, D.~L., {Chatterjee}, S., {Gaensler}, B.~M., \& {Anderson}, J. 2008,
  \apj, 677, 1201

\bibitem[{{Kennel} \& {Coroniti}(1984{\natexlab{a}})}]{kennel84}
{Kennel}, C.~F. \& {Coroniti}, F.~V. 1984{\natexlab{a}}, ApJ, 283, 694

\bibitem[{{Kennel} \& {Coroniti}(1984{\natexlab{b}})}]{kennel84b}
---. 1984{\natexlab{b}}, \apj, 283, 710

\bibitem[{Kothes {et~al.}(2008)Kothes, Landecker, Reich, Safi-Harb, , \&
  Arzoumanian}]{kothes08}
Kothes, R., Landecker, T.~L., Reich, W., Safi-Harb, S., , \& Arzoumanian, Z.
  2008, \apj, 687, 516

\bibitem[{{Kruskal} \& {Schwarzschild}(1954)}]{kruskal54}
{Kruskal}, M. \& {Schwarzschild}, M. 1954, Royal Society of London Proceedings
  Series A, 223, 348

\bibitem[{{LaMassa} {et~al.}(2008){LaMassa}, {Slane}, \& {de
  Jager}}]{lamassa08}
{LaMassa}, S.~M., {Slane}, P.~O., \& {de Jager}, O.~C. 2008, \apjl, 689, L121

\bibitem[{{Malyshev} {et~al.}(2009){Malyshev}, {Cholis}, \&
  {Gelfand}}]{malyshev09}
{Malyshev}, D., {Cholis}, I., \& {Gelfand}, J. 2009, ArXiv e-prints

\bibitem[{{Milne}(1980)}]{milne80}
{Milne}, D.~K. 1980, \aap, 81, 293

\bibitem[{{Mori} {et~al.}(2004){Mori}, {Burrows}, {Hester}, {Pavlov},
  {Shibata}, \& {Tsunemi}}]{mori04}
{Mori}, K., {Burrows}, D.~N., {Hester}, J.~J., {Pavlov}, G.~G., {Shibata}, S.,
  \& {Tsunemi}, H. 2004, \apj, 609, 186

\bibitem[{{Ostriker} \& {Gunn}(1971)}]{ostriker71}
{Ostriker}, J.~P. \& {Gunn}, J.~E. 1971, \apjl, 164, L95+

\bibitem[{{Pacholczyk}(1970)}]{pacholczyk70}
{Pacholczyk}, A.~G. 1970, {Radio astrophysics. Nonthermal processes in galactic
  and extragalactic sources} (Series of Books in Astronomy and Astrophysics,
  San Francisco: Freeman, 1970)

\bibitem[{{Pacini} \& {Salvati}(1973)}]{pacini73}
{Pacini}, F. \& {Salvati}, M. 1973, \apj, 186, 249

\bibitem[{{Rees} \& {Gunn}(1974)}]{rees74}
{Rees}, M.~J. \& {Gunn}, J.~E. 1974, MNRAS, 167, 1

\bibitem[{{Reynolds} \& {Chevalier}(1984)}]{reynolds84}
{Reynolds}, S.~P. \& {Chevalier}, R.~A. 1984, \apj, 278, 630

\bibitem[{{Rybicki} \& {Lightman}(1979)}]{rybicki79}
{Rybicki}, G.~B. \& {Lightman}, A.~P. 1979, {Radiative processes in
  astrophysics} (New York, Wiley-Interscience, 1979.~393 p.)

\bibitem[{{Sedov}(1959)}]{sedov59}
{Sedov}, L.~I. 1959, {Similarity and Dimensional Methods in Mechanics}
  (Similarity and Dimensional Methods in Mechanics, New York: Academic Press,
  1959)

\bibitem[{{Sefako} \& {de Jager}(2003)}]{sefako03}
{Sefako}, R.~R. \& {de Jager}, O.~C. 2003, \apj, 593, 1013

\bibitem[{{Shapiro} \& {Teukolsky}(1986)}]{shapiro86}
{Shapiro}, S.~L. \& {Teukolsky}, S.~A. 1986, {Black Holes, White Dwarfs and
  Neutron Stars: The Physics of Compact Objects} (Black Holes, White Dwarfs and
  Neutron Stars: The Physics of Compact Objects, by Stuart L.~Shapiro, Saul
  A.~Teukolsky, pp.~672.~ISBN 0-471-87316-0.~Wiley-VCH , June 1986.)

\bibitem[{{Slane} {et~al.}(2008){Slane}, {Helfand}, {Reynolds}, {Gaensler},
  {Lemiere}, \& {Wang}}]{slane08}
{Slane}, P., {Helfand}, D.~J., {Reynolds}, S.~P., {Gaensler}, B.~M., {Lemiere},
  A., \& {Wang}, Z. 2008, \apjl, 676, L33

\bibitem[{{Slane} {et~al.}(2004){Slane}, {Helfand}, {van der Swaluw}, \&
  {Murray}}]{slane04}
{Slane}, P., {Helfand}, D.~J., {van der Swaluw}, E., \& {Murray}, S.~S. 2004,
  \apj, 616, 403

\bibitem[{{Spitkovsky}(2008)}]{spitkovsky08}
{Spitkovsky}, A. 2008, \apjl, 682, L5

\bibitem[{{Stone} \& {Gardiner}(2007)}]{stone07}
{Stone}, J.~M. \& {Gardiner}, T. 2007, \apj, 671, 1726

\bibitem[{{Taylor}(1950)}]{taylor50}
{Taylor}, G. 1950, Royal Society of London Proceedings Series A, 201, 159

\bibitem[{{Temim} {et~al.}(2006){Temim}, {Gehrz}, {Woodward}, {Roellig},
  {Smith}, {Rudnick}, {Polomski}, {Davidson}, {Yuen}, \& {Onaka}}]{temim06}
{Temim} et al. 2006, \aj, 132, 1610

\bibitem[{{Temim} {et~al.}(2009){Temim}, {Slane}, {Gaensler}, {Hughes}, \& {Van
  Der Swaluw}}]{temim09}
{Temim}, T., {Slane}, P., {Gaensler}, B.~M., {Hughes}, J.~P., \& {Van Der
  Swaluw}, E. 2009, \apj, 691, 895

\bibitem[{{Truelove} \& {McKee}(1999)}]{truelove99}
{Truelove}, J.~K. \& {McKee}, C.~F. 1999, \apjs, 120, 299

\bibitem[{{van der Swaluw}(2003)}]{vdswaluw03}
{van der Swaluw}, E. 2003, \aap, 404, 939

\bibitem[{{van der Swaluw} {et~al.}(2001){van der Swaluw}, {Achterberg},
  {Gallant}, \& {T{\'o}th}}]{vdswaluw01b}
{van der Swaluw}, E., {Achterberg}, A., {Gallant}, Y.~A., \& {T{\'o}th}, G.
  2001, \aap, 380, 309

\bibitem[{{van der Swaluw} {et~al.}(2004){van der Swaluw}, {Downes}, \&
  {Keegan}}]{vdswaluw04}
{van der Swaluw}, E., {Downes}, T.~P., \& {Keegan}, R. 2004, \aap, 420, 937

\bibitem[{{van der Swaluw} \& {Wu}(2001)}]{vdswaluw01}
{van der Swaluw}, E. \& {Wu}, Y. 2001, \apjl, 555, L49

\bibitem[{{Venter} \& {de Jager}(2006)}]{venter06}
{Venter}, C. \& {de Jager}, O.~C. 2006, ArXiv Astrophysics e-prints (astro-ph/0612652) 

\bibitem[{{Volpi} {et~al.}(2008){Volpi}, {Del Zanna}, {Amato}, \&
  {Bucciantini}}]{volpi08}
{Volpi}, D., {Del Zanna}, L., {Amato}, E., \& {Bucciantini}, N. 2008, \aap,
  485, 337

\bibitem[{{Weisskopf} {et~al.}(2000){Weisskopf}, {Hester}, {Tennant}, {Elsner},
  {Schulz}, {Marshall}, {Karovska}, {Nichols}, {Swartz}, {Kolodziejczak}, \&
  {O'Dell}}]{weisskopf00}
{Weisskopf}, M.~C. et al. 2000, \apjl, 536,  L81

\bibitem[{{Woltjer} {et~al.}(1997){Woltjer}, {Salvati}, {Pacini}, \&
  {Bandiera}}]{woltjer97}
{Woltjer}, L., {Salvati}, M., {Pacini}, F., \& {Bandiera}, R. 1997, \aap, 325,
  295

\bibitem[{{Zwicky}(1938)}]{zwicky38}
{Zwicky}, F. 1938, \apj, 88, 522

\end{thebibliography}

\onecolumn
\appendix
\section{Evolution of a Non-Radiative Supernova Remnant}
\label{snrtheory}

In this model, we assume that the progenitor core-collapse supernova
injected material with mass $M_{\rm ej}$ and initial kinetic energy
$E_{\rm sn}$ into a constant density $\rho_{\rm ISM}$ interstellar
medium (ISM) environment.  We assume the supernova ejecta initially
comprise a constant density inner core surrounded by a $\rho \propto
r^{-9}$ outer envelope -- the standard assumption for core-collapse
supernovae (e.g. \citealt{blondin01}) -- where the transition velocity
$v_t$ between these two components is (Equation 3 in
\citealt{blondin01}):
\begin{eqnarray}
\label{vteqn}
v_t & = & \left(\frac{40E_{\rm sn}}{18M_{\rm ej}} \right)^{1/2}.
\end{eqnarray}
Under this assumption, the density of material inside the SNR evolves
as (Equations 1 \& 2 in \citealt{blondin01}):
\begin{eqnarray}
\label{rhosnr}
\rho_{\rm ej}(r,t) & = & \left\{\begin{array}{ll}
\frac{10}{9\pi}E_{\rm sn}v_t^{-5} t^{-3} & r<v_t t \\
\frac{10}{9\pi}E_{\rm sn}v_t^{-5}\left(\frac{r}{v_t t}\right)^{-9}t^{-3} &
r>v_t t
\end{array} \right. .
\end{eqnarray}
Initially, the evolution of the SNR is self-similar
(e.g. \citealt{chevalier82}), and therefore described in terms of
characteristic length ($R_{\rm ch}$), time ($t_{\rm ch}$), and mass
($M_{\rm ch}$) scales.  For a SNR expanding into a constant density
ISM, these scales are (Equations 1--3 in \citealt{truelove99}):
\begin{eqnarray}
\label{rsnrchar}
R_{\rm ch} & \equiv & M_{\rm ej}^{1/3} \rho_{\rm ISM}^{-1/3}, \\
\label{tsnrchar}
t_{\rm ch} & \equiv & E_{\rm sn}^{-1/2} M_{\rm ej}^{5/6} \rho_{\rm
  ISM}^{-1/3}, {\rm and} \\
\label{msnrchar}
M_{\rm ch} & \equiv & M_{\rm ej} .
\end{eqnarray}
The expanding supernova ejecta drives a shock wave into the
surrounding ISM (called the ``forward shock'') which marks the outer
boundary of the SNR ($R_{\rm snr}$).  Initially the dynamics of the
SNR is dominated by expanding supernova ejecta because the ejecta mass
is much greater than the mass of the ISM material swept up and shocked
by the SNR ($M_{\rm sw,snr} \equiv \frac{4}{3}\pi R_{\rm snr}^3
\rho_{\rm ISM}$).  During this period, $R_{\rm snr}(t)$ is (Equation
75 in \citealt{truelove99}):
\begin{eqnarray}
\label{rsnr1}
R_{\rm snr}(t) & = & 1.12 R_{\rm ch} \left(\frac{t}{t_{\rm ch}} \right)^{2/3} ,
\end{eqnarray}
where 1.12 is specific for a $\rho \propto r^{-9}$ outer ejecta
envelope, though it varies little for different values of the
power-law exponent \citep{truelove99}.  Therefore, the expansion
velocity $v_{\rm snr}$ ($v_{\rm snr}(t) \equiv \frac{dR_{\rm
snr}}{dt}$) of the SNR is (Equation 76 in \citealt{truelove99}):
\begin{eqnarray}
\label{vsnr1}
v_{\rm snr} = 0.75 \frac{R_{\rm ch}}{t_{\rm ch}} \left(\frac{t}{t_{\rm
ch}} \right)^{-1/3}.
\end{eqnarray}
As the SNR grows, $M_{\rm sw,snr}$ increases and eventually will reach
the point where $M_{\rm sw,snr} \approx M_{\rm ej}$.  At this point,
the swept-up ISM material will begin to dominate the dynamics of the
SNR, and the SNR is said to enter the Sedov-Taylor phase of its
evolution \citep{sedov59, taylor50}.  For a SNR with the ejecta
profile given in Equation \ref{rhosnr}, this transition occurs at time
$t_{\rm ST} \simeq 0.52t_{\rm ch}$ \citep{truelove99}.  During the
Sedov-Taylor phase, $R_{\rm snr}$ is (Equation 56 in
\citealt{truelove99}):
\begin{eqnarray}
\label{rsnr2}
R_{\rm snr}(t) & = & \left[R_{\rm snr,ST}^{5/2} + \left(2.026
    \frac{E_{\rm sn}}{\rho_{\rm ISM}}\right)^{1/2}(t-t_{\rm ST}) \right]^{2/5},
\end{eqnarray}
where $R_{\rm snr,ST}\equiv R_{\rm snr}(t_{\rm ST})$.  The
Sedov-Taylor phase ends when the gas recently shocked by the expanding
SNR cools radiatively, which occurs approximately at (Equation 3 in
\citealt{blondin98}):
\begin{eqnarray}
\label{trad}
t_{\rm rad} & = & 2.9 \left(\frac{E_{\rm sn}}{10^{51}~{\rm
    ergs}}\right)^{4/17} \left( \frac{\rho_{\rm ism}}{m_{p}~{\rm cm}^{-3}}
    \right)^{-9/17}\times10^{4}~{\rm years}
\end{eqnarray}
after the supernova, where $m_{p}$ is the mass of a proton.

The pressure $P$ and density $\rho$ profile of a SNR changes
significantly over time.  Initially, the expansion velocity of the SNR
is significantly larger than the sound speed of the surrounding ISM.
As a result, the swept-up material is shocked by the surrounding
ejecta.  Assuming that energy losses due to cosmic ray acceleration is
negligible, the pressure of the recently shocked ISM material [$P_{\rm
snr}(R_{\rm snr})$] is:
\begin{eqnarray}
\label{psnr}
P_{\rm snr}(R_{\rm snr},t) & = & \frac{3}{4} \rho_{\rm ism} v_{\rm
  snr}(t)^2,
\end{eqnarray}
assuming an adiabatic index $\gamma=5/3$ (non-relativistic gas).  The
pressure of the shocked ISM is significantly higher than that of the
ejecta inside the SNR, which due to adiabatic expansion has cooled
significantly since the explosion.  As a result, the layer of shocked
ISM expands inside the SNR, driving a shock wave, referred to as the
reverse shock, into the supernova ejecta.  When the reverse shock is
in the outer envelope of the supernova ejecta, its radius $R_{\rm rs}$
is \citep{chevalier82,truelove99}:
\begin{eqnarray}
\label{radrs1}
R_{\rm rs}(t) & = & \frac{1}{1.19}R_{\rm snr}(t),
\end{eqnarray}
and therefore, its velocity relative to the surrounding ISM $v_{\rm
rs}$ is:
\begin{eqnarray}
\label{velrs1}
v_{\rm rs}(t) & = & \frac{1}{1.19}v_{\rm snr}(t).
\end{eqnarray}
The properties of ejecta recently shocked by the reverse shock depends
on the velocity of the reverse shock relative to the unshocked ejecta,
$v_{\rm ej}$.  The standard assumption is the unshocked ejecta is
expanding ballisticly [$v_{\rm ej}(r,t) \equiv r/t$].  Therefore, the
relative velocity of the reverse shock, $\tilde{v}_{\rm rs}\equiv
v_{ej}(R_{\rm rs},t) - v_{\rm rs}(t)$, is \citep{truelove99}:
\begin{eqnarray}
\tilde{v}_{\rm rs}(t) & = & \frac{1}{2.38}v_{\rm snr}(t).
\end{eqnarray}
Eventually, the reverse shock will enter the constant density core at
the center of the SNR.  For a SNR with an initial ejecta density
profile defined in Equation (\ref{rhosnr}), this occurs a time $t_{\rm
core} \simeq 0.25t_{\rm ch}$ (Equation 79 in \citealt{truelove99})
where $t_{\rm ch}$ is defined above in Equation (\ref{tsnrchar}).
After this time, the radius of the reverse shock evolves as (Equation
83 in \citealt{truelove99}):
\begin{eqnarray}
\label{radrs2}
R_{\rm rs}(t) & = & \left[1.49 - 0.16\frac{t - t_{\rm core}}{t_{\rm
      ch}} - 0.46\ln\left(\frac{t}{t_{\rm core}}\right) \right]
      \frac{R_{\rm ch}}{t_{\rm ch}}t.
\end{eqnarray}
During this stage, $\tilde{v}_{\rm rs}$ is equal to (Equation 84 in
\citealt{truelove99}): 
\begin{eqnarray}
\tilde{v}_{\rm rs} & = & \left[0.50+0.16\frac{(t-t_{\rm core})}{t_{\rm
      ch}} \right]\frac{R_{\rm ch}}{t_{\rm ch}}.
\end{eqnarray}

Since the velocity of the reverse shock is much greater than the sound
speed in the unshocked ejecta, the reverse shock -- like the forward
shock -- is a strong shock.  Therefore, the pressure of ejecta
recently shocked by the reverse shock $P_{\rm rs}$, is \citep{truelove99}:
\begin{eqnarray}
P_{\rm rs}(t) & = & \frac{\rho_{\rm ej}(R_{\rm rs},t)\tilde{v}_{\rm
    rs}(t)^2} {\rho_{\rm ISM}v_{\rm snr}(t)^2} P_{\rm snr}(t).
\end{eqnarray} 
When the reverse shock is in the outer envelope of the supernova
ejecta, the density, velocity, and pressure profile of the SNR between
the reverse shock and forward shock ($R_{\rm rs}<r<R_{\rm snr}$) is
calculated using the equations derived by \citet{chevalier82}.  These
equations are no longer valid once the reverse shock enters the
constant density ejecta core.  At this point, we model the density,
velocity, and pressure structure of the SNR using the solution for a
Sedov-Taylor SNR presented in Appendix A of \citet{bandiera84}.  This
is valid until the SNR goes radiative ($t=t_{\rm rad}$).
Unfortunately, the internal structure of a radiative SNR is poorly
understood, and therefore we do not attempt to model the evolution of
a PWN in this environment.  However, previous work suggests that the
radius of a PWN $R_{\rm pwn}$ inside a radiative SNR evolves as
\citep{blondin01,vdswaluw01}:
\begin{equation}
\frac{R_{\rm pwn}}{R_{\rm snr}} \propto t^{0.075}.
\end{equation}

\section{Initial Properties of a PWN inside a SNR}
\label{initialcond}
In order to model the evolution of a PWN inside a SNR, it is necessary
to estimate the initial conditions.  To estimate the initial energy of
the PWN, we assume that adiabatic losses dominate and the PWN is
expanding with a constant velocity.  In this case:
\begin{equation}
\frac{dE_{\rm pwn}}{dt} = \dot{E_0} \left(1+\frac{t}{\tau_{\rm
sd}}\right)^{-\frac{p+1}{p-1}} - \frac{E_{\rm pwn}}{t}.
\end{equation}
Defining $y=(p+1)/(p-1)$, $\epsilon \equiv E_{\rm pwn} /
(\dot{E_0}\tau_{\rm sd})$, and $x \equiv t/\tau_{\rm sd}$, one derives
that:
\begin{equation}
\frac{d\epsilon}{dx} = -\frac{\epsilon}{x} + (1+x)^{-y}.
\end{equation}
Using the boundary conditions that $\epsilon(x=0)=0$ (initially, the
PWN has zero energy), one derives that, for $y=2$ ($p=3$):
\begin{equation}
\epsilon=\frac{\ln(1+x)}{x}-\frac{1}{x+1}
\end{equation}
and, for $y\neq2$ ($p\neq3$),
\begin{equation}
\label{epwninit}
\epsilon=\frac{(1+x)^{1-y}}{1-y} - \frac{(1+x)^{2-y}}{x(1-y)(2-y)} +
\frac{1}{x(1-y)(2-y)}.
\end{equation}
We also assume that initially the fraction of the PWN's energy in
magnetic fields is $\eta_B$, the fraction of the PWN's energy in
electrons and positrons is $\eta_e$, and the fraction of the PWN's
energy in ions is $\eta_i$.  Additionally, we assume that initial
spectrum of particles inside the PWN has the same shape as the
injection spectrum.  

If adiabatic losses dominate, the equation of motion of the PWN for
$P_{\rm snr}(R_{\rm pwn})\equiv0$ is \citep{ostriker71, chevalier92}:
\begin{equation} 
\label{rpwninit1}
M_{\rm sw,pwn} \frac{dv_{\rm pwn}}{dt} = 4\pi R_{\rm pwn}^2 [P_{\rm
    pwn} - \rho_{\rm ej}(R_{\rm pwn}) \times (v_{\rm pwn}-v_{\rm
    ej}(R_{\rm pwn}))^2 ],
\end{equation}
where:
\begin{equation}
\label{rpwninit2}
\frac{dE_{\rm pwn}}{dt} = \dot{E} - P_{\rm pwn} 4\pi R_{\rm pwn}^2
v_{\rm pwn}.
\end{equation}
At early times ($t\ll\tau$), $\dot{E}\approx\dot{E}_0$, and it is
possible to solve Equations \ref{rpwninit1} \& \ref{rpwninit2}
analytically (e.g. \citealt{chevalier92}).  For the initial supernova
ejecta density given in Equation \ref{rhosnr}, $R_{\rm pwn}$ can be
expressed as (e.g. Equation 2.6 in \citealt{chevalier92}, Equation 5
in \citealt{blondin01}):
\begin{equation}
\label{rpwninit}
R_{\rm pwn}(t) = 1.44 \left(\frac{E_{\rm sn}^3\dot{E}_0^2}{M_{\rm
    ej}^5} \right)^{1/10} t^{6/5}.
\end{equation}

\end{document}